\documentclass[aps,prl,preprint,groupedaddressm,showkeys]{revtex4-1}
\usepackage{graphicx} 
\usepackage{epstopdf} 
\usepackage{setspace}
\usepackage{rotating}
\usepackage{multirow}
\usepackage{amsmath,amsthm}

\begin{document}


\title{Tracking Employment Shocks Using Mobile Phone Data}


\author{Jameson L. Toole}
\email[]{jltoole@mit.edu}
\affiliation{Engineering Systems Division, MIT, Cambridge, MA, 02144}

\author{Yu-Ru Lin}
\affiliation{School of Information Science, University of Pittsburgh, Pittsburgh, PA 15260}

\author{Erich Muehlegger}
\affiliation{Department of Economics, UC Davis, Davis, CA 95616}

\author{Daniel Shoag}
\affiliation{Harvard Kennedy School, Harvard University, Cambridge, MA, 02144}

\author{Marta C. Gonz\'alez}
\affiliation{Department of Civil and Environmental Engineering, MIT, Cambridge, MA, 02144}

\author{David Lazer}
\affiliation{Lazer Laboratory, Department of Political Science and College of Computer and Information Science, Northeastern University, Boston, MA 02115, USA}
\affiliation{Harvard Kennedy School, Harvard University, Cambridge, MA, 02144}


\begin{abstract}
Can data from mobile phones be used to observe economic shocks and their consequences at multiple scales? Here we present novel methods to detect mass layoffs, identify individuals affected by them, and predict changes in aggregate unemployment rates using call detail records (CDRs) from mobile phones. Using the closure of a large manufacturing plant as a case study, we first describe a structural break model to correctly detect the date of a mass layoff and estimate its size.  We then use a Bayesian classification model to identify affected individuals by observing changes in calling behavior following the plant's closure.  For these affected individuals, we observe significant declines in social behavior and mobility following job loss. Using the features identified at the micro level, we show that the same changes in these calling behaviors, aggregated at the regional level, can improve forecasts of macro unemployment rates. These methods and results highlight promise of new data resources to measure micro economic behavior and improve estimates of critical economic indicators.
\end{abstract}

\maketitle
Economic statistics are critical for decision-making by both government and private institutions. Despite their great importance, current measurements draw on limited sources of information, losing precision with potentially dire consequences. The beginning of the Great Recession offers a powerful case study: the initial BEA estimate of the contraction of GDP in the fourth quarter of 2008 was an annual rate 3.8\%. The American Recovery and Reinvestment Act (stimulus) was passed based on this understanding in February 2009. Less than two weeks after the plan was passed, that 3.8\% figure was revised to 6.2\%, and subsequent revisions peg the number at a jaw dropping 8.9\% -- more severe than the worst quarter during the Great Depression. The government statistics were wrong and may have hampered an effective intervention.  As participation rates in unemployment surveys drop, serious questions have been raised as to the declining accuracy and increased bias in unemployment numbers~\cite{Krueger2014}. 

\begin{figure}
 \includegraphics[width=0.5\linewidth]{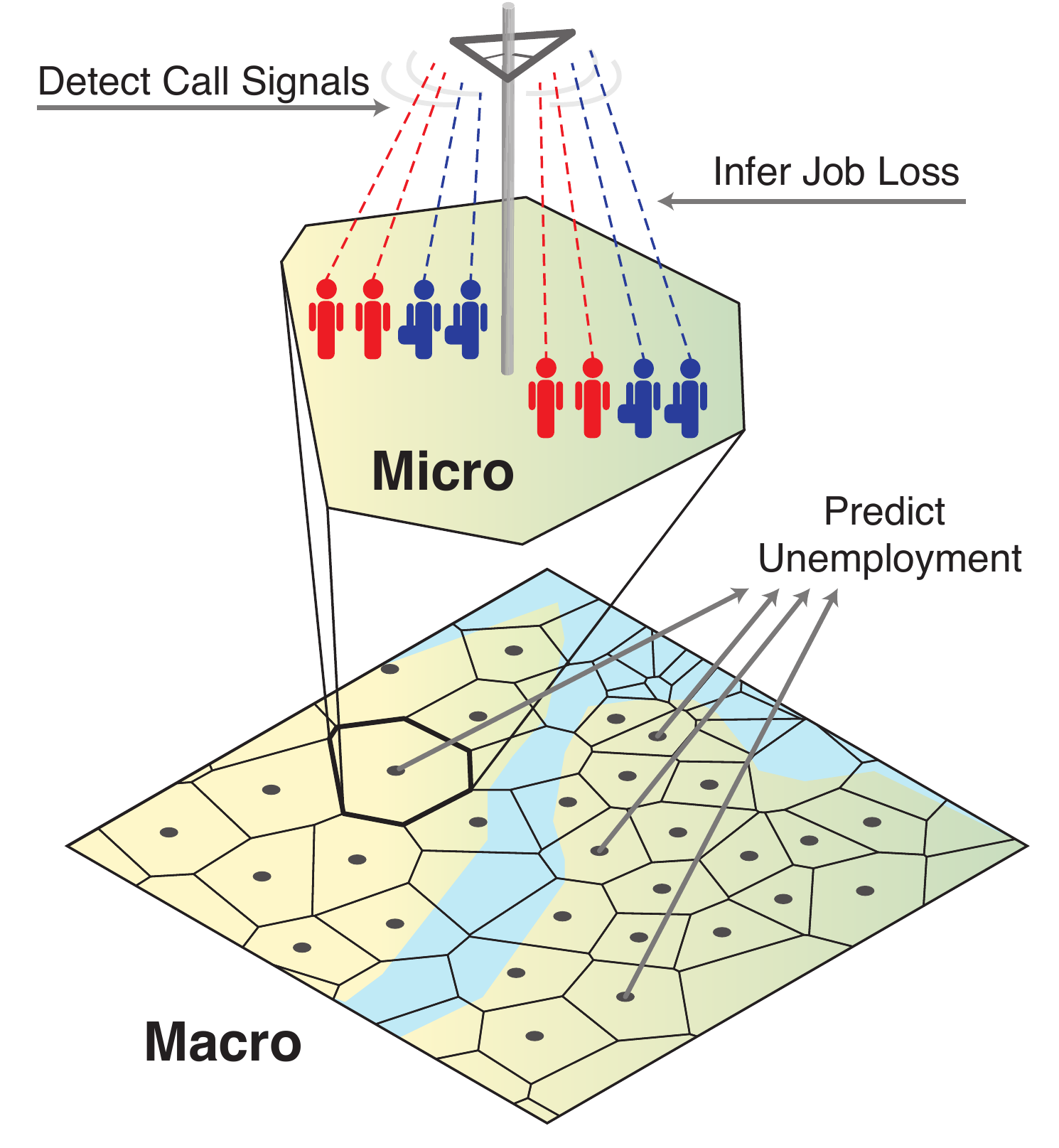}%
 \caption{\label{fig:schematic}A schematic view of the relationship between job loss and call dynamics.  We use the calling behavior of individuals to infer job loss and measure its effects.  We then measure these variables and include them in predictions of unemployment at the macro scale, significantly improving forecasts.}
 \end{figure}

In this paper we offer a methodology to infer changes in the macro economy in near real time, at arbitrarily fine spatial granularity, using data already passively collected from mobile phones. We demonstrate the reliability of these techniques by studying data from two European countries. In the first, we show it is possible to observe mass layoffs and identify the users affected by them in mobile phone records. We then track the mobility and social interactions of these affected workers and observe that job loss has a systematic dampening effect on their social and mobility behavior. Having observed an effect in the micro data, we apply our findings to the macro scale by creating corresponding features to predict unemployment rates at the province scale. In the second country, where the macro-level data is available, we show that changes in mobility and social behavior predict unemployment rates ahead of official reports and more accurately than traditional forecasts.  These results demonstrate the promise of using new data to bridge the gap between micro and macro economic behaviors and track important economic indicators. Figure \ref{fig:schematic} shows a schematic of our methodology.

\clearpage

\section{Measuring the Economy}
Contemporary macroeconomic statistics are based on a paradigm of data collection and analysis begun in the 1930s \cite{Marcuss2007, Card2011}. Most economic statistics are constructed from either survey data or administrative records. For example, the US unemployment rate is calculated based on the monthly Current Population Survey of roughly 60,000 households, and the Bureau of Labor Statistics manually collects 80,000 prices a month to calculate inflation. Both administrative databases and surveys can be slow to collect, costly to administer, and fail to capture significant segments of the economy.  These surveys can quickly face sample size limitations at fine geographies and require strong assumptions about the consistency of responses over time.  Statistics inferred from survey methods have considerable uncertainty and are routinely revised in months following their release as other data is slowly collected \cite{Hausman2004, Jones1999, Burda2012, Krueger2014}. Moreover, changes in survey methodology can result in adjustments of reported rates of up to 1-2 percentage points~\cite{Tiller1994}.

The current survey-based paradigm also makes it challenging to study the effect of economic shocks on networks or behavior without reliable self-reports. This has hampered scientific research. For example, many studies have documented the severe negative consequences of job loss in the form of difficulties in retirement~\cite{Chan1999}, persistently lower wages following re-employment including even negative effects on children's outcomes~\cite{Ruhm1999, Oreopoulos2008}, increased risk of death and illness~\cite{Sullivan2009, Classen2012}, higher likelihood of divorce~\cite{Charles2001}, and, unsurprisingly, negative impacts and on happiness and emotional well-being~\cite{Krueger2011}.  Due to the cost of obtaining the necessary data, however, social scientists have been unable to directly observe the large negative impact of a layoff on the frequency and stability of an individual's social interactions or mobility.

\section{Predicting the Present}
These shortcomings raise the question as to whether existing methods could be supplemented by large-scale behavioral trace data. There have been substantial efforts to discern important population events from such data, captured by the pithy phrase of, ``predicting the present" \cite{Choi2009, Vespignani2009, Lazer2009, Henderson2012}. Prior work has linked news stories with stock prices \cite{Hayo2005, Lavrenko2000, Schumaker2009} and used web search or social media data to forecast labor markets \cite{Ettredge2005, Askitas2009, Eagle2006, Suhoy2009, Antenucci2014}, consumer behavior \cite{Goel2010, Gelman2014}, automobile demand, vacation destinations \cite{Choi2009, Choi2012}.  Research on social media, search, and surfing behavior have been shown to signal emerging public health problems \cite{Ginsberg2008, Aramaki2011, Chew2010, Culotta2010, Gomide2011, Quincey2010, Zamite2011, Signorini2011}; although for a cautionary tale see \cite{Lazer2014}. And recent efforts have even been made towards leveraging Twitter to detect and track earthquakes in real-time detection faster than seismographic sensors \cite{Sakaki2010, Guy2010, Earle2012}. While there are nuances to the analytic approaches taken, the dominant approach has been to extract features from some large scale observational data and to evaluate the predictive (correlation) value of those features with some set of measured aggregate outcomes (such as disease prevalence).  Here we offer a twist on this methodology through identification of features from observational data and to cross validate across individual and aggregate levels.

All of the applications of predicting the future are predicated in part on the presence of distinct signatures associated with the systemic event under examination.  The key analytic challenge is to identify signals that (1) are observable or distinctive enough to rise above the background din, (2) are unique or generate few false positives, (3) contain information beyond well-understood patterns such as calendar-based fluctuations, and (4) are robust to manipulation. Mobile phone data, our focus here, are particularly promising for early detection of systemic events as they combine spatial and temporal comprehensiveness, naturally incorporate mobility and social network information, and are too costly to intentionally manipulate.

Data from mobile phones has already proven extremely beneficial to understanding the everyday dynamics of social networks \cite{Barabasi2005, Rybski2009, Onnela2011, Onnela2007, Cho2011, Goncalves2008, Bond2012} and mobility patterns of millions \cite{Balcan2009, Han2011, Brockmann2006, Gonzalez2008, Song2010a, Song2010b, Deville2014, Yan2014}.  With a fundamental understanding of regular behavior, it becomes possible to explore deviations caused by collective events such as emergencies \cite{Bagrow2011}, natural disasters \cite{Blumenstock2011, Lu2012}, and cultural occasions \cite{Calabrese2011, Calabrese2006}.  Less has been done to link these data to economic behavior. In this paper we offer a methodology to robustly infer changes to measure employment shocks at extremely high spatial and temporal resolutions and improve critical economic indicators.

\section{Data}
We focus our analysis at three levels:  the individual, the community, and the provincial levels.  We begin with unemployment at the community (town) level, where we examine the behavioral traces of a large-scale layoff event.  At the community and individual levels, we analyze call record data from a service provider with an approximately 15\% market share in an undisclosed European country.  The community-level data set spans a 15 month period between 2006 and 2007, with the exception of a 6 week gap due to data extraction failures.  At the province level, we examine call detail records from a service provider from another European country, with an approximately 20\% market share and data running for 36 months from 2006 to 2009. Records in each data set include an anonymous id for caller and callee, the location of the tower through which the call was made, and the time the call occurred.  In both cases we examine the universe of call records made over the provider's network (see SI for more details).

\section{Observing Unemployment at the Community Level}
We study the closure of an auto-parts manufacturing plant (the plant) that occurred in December, 2006.  As a result of the plant closure, roughly 1,100 workers lost their jobs in a small community (the town) of 15,000. Our approach builds on recent papers \cite{Gonzalez2008, Song2010a, Song2010b, Bagrow2011} that use call record data to measure social and mobility patterns. 

There are three mobile phone towers within close proximity of the town and the plant. The first is directly within the town, the second is roughly 3$km$ from the first and is geographically closest to the manufacturing plant, while the third is roughly 6.5$km$ from the first two on a nearby hilltop.  In total, these three towers serve an area of roughly 220$km^2$ of which only 6$km^2$ is densely populated. There are no other towns in the region covered by these towers. Because the exact tower through which a call is routed may depend on factors beyond simple geographic proximity (e.g. obstructions due to buildings), we consider any call made from these three towers as having originated from the town or plant.

We model the pre-closure daily population of the town as made up of a fraction of individuals $\gamma$ who will no longer make calls near the plant following its closure and the complimentary set of individuals who will remain $(1-\gamma)$.  As a result of the layoff, the total number of calls made near the plant will drop by an amount corresponding to the daily calls of workers who are now absent.  This amounts to a structural break model that we can use to estimate the prior probability that a user observed near the plant was laid off, the expected drop in calls that would identify them as an affected worker, and the time of the closure (see SI for full description of this model). We suspect that some workers laid off from the plant are residents of the town and thus they will still appear to make regular phone calls from the three towers and will not be counted as affected.  Even with this limitation, we find a large change in behavior.

\begin{figure}[h]
 \includegraphics[width=1\linewidth]{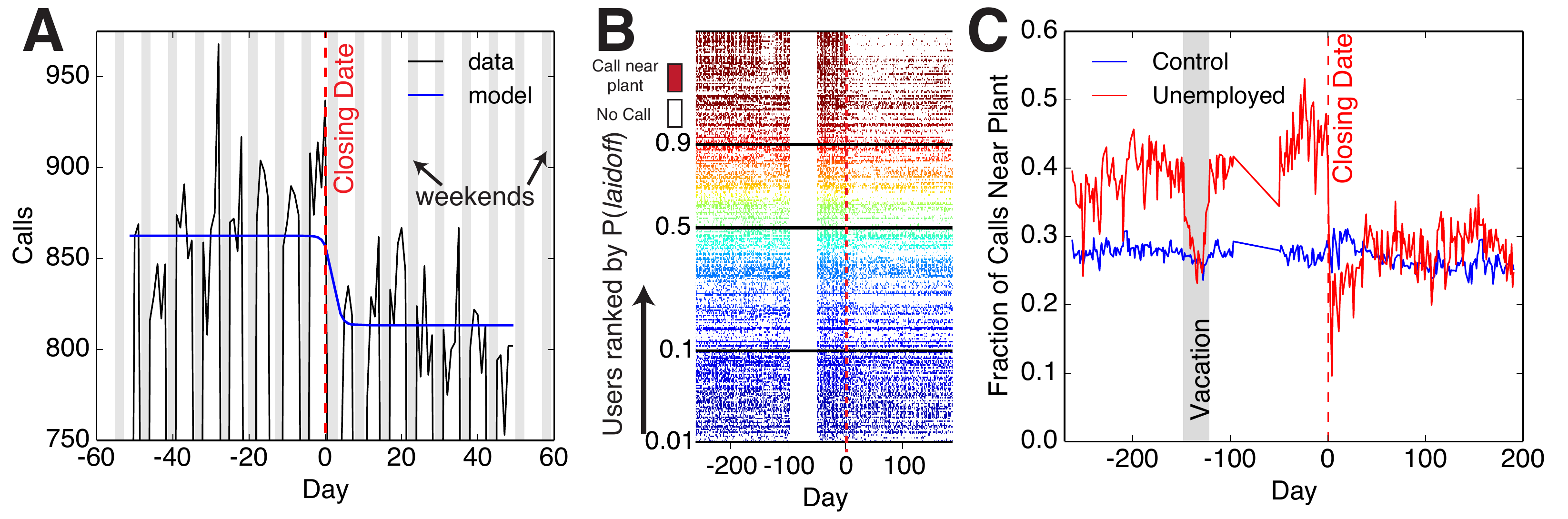}%
 \caption{\label{fig:layoff_date}Identifying the layoff date. A) Total aggregate call volume (black line) from users who make regular calls from towers near the plant is plotted against our model (blue).  The model predicts a sudden drop in aggregate call volume and correctly identifies the date of the plant closure as the one reported in newspapers and court records. B) Each of the top 300 users likely to have been laid off is represented by a row where we fill in a day as colored if a call was made near the plant on that day.  White space marks the absence of calls. Rows are sorted by the assigned probability of that user being laid off according to our Bayesian model.  Users with high probabilities cease making calls near the plant directly following the layoff. C) We see a sharp, sustained drop in the fraction of calls made near the plant by users assigned to the top decile in probability of being unemployed (red) while no affect is seen for the control group users believed to be unaffected (blue).  Moreover, we see that laid off individuals have an additional drop off for a two week period roughly 125 days prior the plant closure.  This time period was confirmed to be a coordinated vacation for workers providing further evidence we are correctly identifying laid off workers.}
 \end{figure}

To verify the date of the plant closing, we sum the number of daily calls from 1955 regular users (i.e. those who make at least one call from the town each month prior to the layoff) connecting through towers geographically proximate to the affected plant.  The estimator selects a break date, $t_{layoff}$ , and pre- and post- break daily volume predictions to minimize the squared deviation of the model from the data. The estimated values are overlaid on daily call volume and the actual closure date in the Figure \ref{fig:layoff_date}A. As is evident in the figure, the timing of the plant closure (as reported in newspapers and court filings) can be recovered statistically using this procedure - the optimized predictions display a sharp and significant reduction at this date. As a separate check to ensure this method is correctly identifying the break date, we estimate the same model for calls from each individual user $i$ and find a distribution of these dates $t_{layoff}^i$ is peaked around the actual layoff date (see Figure 1 in SI).

\section{Observing Unemployment at the Individual Level}
To identify users directly affected by the layoff, we calculate Bayesian probability weights based on changes in mobile phone activity. For each user, we calculate the conditional probability that a user is a non-resident worker laid off as part of the plant closure based on their observed pattern of calls.  To do this, we compute the difference in the fraction of days on which a user made a call near the plant in 50 days prior to the week of the layoff.  We denote this difference as $\Delta q = q_{pre}-q_{post}$.  We consider each user's observed difference a single realization of a random variable, $\Delta q$. Under the hypothesis that there is no change in behavior, the random variable $\Delta q$ is distributed $N(0, \sqrt{\frac{q_{pre}(1-q_{pre})}{50}+\frac{q_{post}(1-q_{post})}{50}})$.  Under the alternative hypothesis the individual's behavior changes pre- and post-layoff, the random variable $\Delta q$ is distributed $N(d, \sqrt{\frac{q_{pre}(1-q_{pre})}{50}+\frac{q_{post}(1-q_{post})}{50}})$, where $d$ is the mean reduction in calls from the plant for non-resident plant workers laid off when the plant was closed.  We assign user $i$ the following probability of having been laid off given his or her calling pattern:

\begin{equation}
P(laidoff)_i = \frac{\gamma P(\Delta \hat{q} | \Delta q =d)}{\gamma P(\Delta \hat{q} | \Delta q = d) + (1-\gamma)P(\Delta \hat{q} | \Delta q=0)}
\end{equation}

Calculating the probabilities requires two parameters, $\gamma$, our prior that an individual is a non-resident worker at the affected plant and $d$, the threshold we use for the alternative hypothesis.  The values of $\gamma=5.8$\% and $d=0.29$ are determined based on values fit from the model in the previous section. 

\subsection{Validating the Layoff}
On an individual level, Figure \ref{fig:layoff_date}B shows days on which each user makes a call near the plant ranked from highest to lowest probability weight (only the top 300 users are shown, see Figure 2 in SI for more users). Users highly suspected of being laid off demonstrate a sharp decline in the number of days they make calls near the plant following the reported closure date.  While we do not have ground-truth evidence that any of these mobile phone users was laid off, we find more support for our hypothesis by examining a two week period roughly 125 days prior to the plant closure.  Figure \ref{fig:layoff_date}C shows a sharp drop in the fraction of calls coming from this plant for users identified as laid off post closure. This period corresponds to a confirmed coordinated holiday for plant workers and statistical analysis confirms a highly significant break for individuals classified as plant workers in the layoff for this period.  Given that we did not use call data from this period in our estimation of the Bayesian model, this provides strong evidence that we are correctly identifying the portion of users who were laid off by this closure. In aggregate, we assign 143 users probability weights between 50\% and 100\%.  This represents 13\% of the pre-closure plant workforce and compares closely with the roughly 15\% national market share of the service provider.

\section{Assessing the Effect of Unemployment at the Individual Level}
\begin{figure}[h]
 \includegraphics[width=1\linewidth]{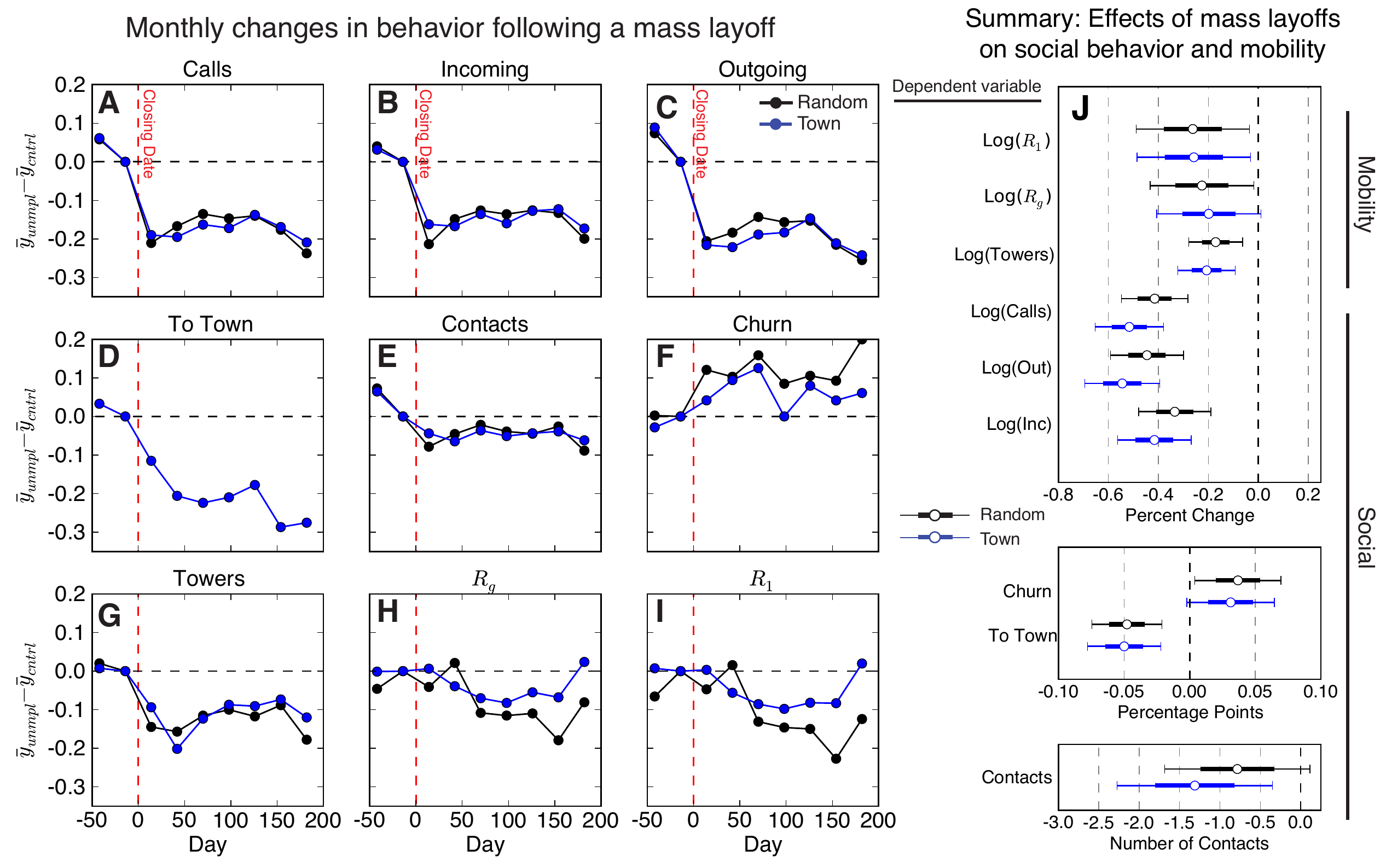}%
 \caption{\label{fig:individual_changes} Changes in social networks and mobility following layoffs. We quantify the effect of mass layoffs relative to two control groups: users making regular calls from the town who were not identified as laid off and a random sample of users from the rest of the country.  We report monthly point estimates for six social and three mobility behaviors: A) Total calls, B) number of incoming calls, C) number of outgoing calls, D) Fraction of calls to individuals in the town at the time of the call, E) number of unique contacts, and the fraction of individuals called in the previous month who were not called in the current month (churn), G) Number of unique towers visited, H) radius of gyration, I) average distance from most visited tower. Pooling months pre- and post-layoff yield statistically significant changes in monthly social network and mobility metrics following a mass layoff. J)  Reports regression coefficient for each of our 9 dependent variables along with the 66\% and 95\% confidence intervals.}
 \end{figure}

We now turn to analyzing behavioral changes associated with job loss at the individual level.  We first consider six quantities related to the monthly social behavior: A) total calls, B) number of incoming calls, C) number of outgoing calls, D) calls made to individuals physically located in the town of the plant (as a proxy for contacts made at work),  E) number of unique contacts, and F) the fraction of contacts called in the previous month that were not called in the current month, referred to as churn.  In addition to measuring social behavior, we also quantify changes in three metrics related to mobility: G) number of unique locations visited, H) radius of gyration, and I) average distance from most visited tower (see SI for detailed definitions of these variables). To guard against outliers such as long trips for vacation or difficulty identifying important locations due to noise, we only consider months for users where more than 5 calls were made and locations where a user recorded more than three calls.

We measure changes in these quantities using all calls made by each user (not just those near the plant) relative to months prior to the plant closure, weighting measurements by the probability an individual is laid off and relative to two reference groups: individuals who make regular calls from the town but were not believed to be laid off (mathematically we weight this group using the inverse weights from our bayesian classifier) and a random sample of 10,000 mobile phone users throughout the country (all users in this sample are weighted equally).

Figure \ref{fig:individual_changes}A-I shows monthly point estimates of the average difference between relevant characteristics of users believed to be laid off compared to control groups. This figure shows an abrupt change in variables in the month directly following the plant closure.  Despite this abrupt change, data at the individual level are sufficiently noisy that the monthly point estimates are not significantly different from 0 in every month.  However, when data from months pre- and post-layoff are pooled, these differences are robust and statistically significant. The right panel of Figure \ref{fig:individual_changes} and Table I in the SI show the results of OLS regressions comparing the pre-closure and post-closure periods for laid-off users relative to the two reference groups (see SI for detailed model specification as well as confidence intervals for percent changes pre- and post-layoff for each variable).  The abrupt and sustained change in monthly behavior of individuals with a high probability of being laid off is compelling evidence in support of using mobile phones to detect mass layoffs with mobile phones.

We find that the total number of calls made by laid off individuals drops 51\% and 41\% following the layoff when compared to non-laid off residents and random users, respectively.  Moreover, this drop is asymmetric. The number of outgoing calls decreases by 54\% percent compared to a 41\% drop in incoming calls (using non-laid off residents as a baseline).  Similarly, the number of unique contacts called in months following the closure is significantly lower for users likely to have been laid off.  The fraction of calls made by a user to someone physically located in the town drops 4.7 percentage points for laid off users compared to residents of the town who were not laid off.  Finally, we find that the month-to-month churn of a laid off person's social network increases roughly 3.6 percentage points relative to control groups.  These results suggest that a user's social interactions  see significant decline and that their networks become less stable following job loss.  This loss of social connections may amplify the negative consequences associated with job loss observed in other studies.

For our mobility metrics, find that the number of unique towers visited by laid-off individuals decreases 17\% and 20\% relative to the random sample and town sample, respectively.  Radius of gyration falls by 20\% and 22\% while the average distance a user is found from the most visited tower also decrease decreases by 26\% relative to reference groups.  These changes reflect a general decline in the mobility of individuals following job loss, another potential factor in long term consequences. 

\section{Observing Unemployment at the Province Level}
 
The relationship between mass layoff events and these features of CDRs suggests a potential for predicting important, large-scale unemployment trends based on the population's call information.  Provided the effects of general layoffs and unemployment are similar enough to those due to mass layoffs, it may be possible to use observed behavioral changes as additional predictors of general levels of unemployment.  To perform this analysis, we use another CDR data set covering approximately 10 million subscribers in a different European country, which has been studied in prior work~\cite{Onnela2011, Onnela2007, Gonzalez2008, Song2010a, Song2010b, Bagrow2011}.  This country experienced enormous macroeconomic disruptions, the magnitude of which varied regionally during the period in which the data are available. We supplement the CDR data set with quarterly, province-level unemployment rates from the EU Labor Force Survey, a large sample survey providing data on regional economic conditions within the EU (see SI for additional details).

We compute seven aggregated measures identified in the previous section: call volume, incoming calls, outgoing calls, number of contacts, churn, number of towers, and radius of gyration.  Distance from home was omitted due to strong correlation with radius of gyration while calls to the town was omitted because it is not applicable in a different country. For reasons of computational cost, we first take a random sample of 3000 mobile phone users for each province. The sample size was determined to ensure the estimation feature values are stable (see SI Figure 6 for details). We then compute the seven features aggregated per month for each individual user. The $k$-th feature value of user $i$ at month $t$ is denoted as $y_{i,t,k}$ and we compute month over month changes in this quantity as $y'_{i,t,k}= \frac{y_{i,t,k}}{y_{i,t-1,k}}$. A normalized feature value for a province $s$, is computed by averaging all users in selected province $\bar{y}_{s,t,k}=\sum_{i\in s}y'_{i,t,k}$. In addition, we use percentiles of the bootstrap distribution to compute the 95\% confidence interval for the estimated feature value. 

After aggregating these metrics to the province level, we assess their power to improve predictions of unemployment rates.  Note that we do not attempt to identify mass layoffs in this country.  Instead, we look for behavioral changes that may have been caused by layoffs and see if these changes are predictive of general unemployment statistics. First, we correlate each aggregate measure with regional unemployment separately, finding significant correlations in the same direction as was found for individuals (see Table II in the SI). We also find the strong correlations between calling behavior variables, suggesting that principal component analysis (PCA) can reasonably be used to construct independent variables that capture changes in calling behavior while guarding against co-linearity.  The first principal component, with an eigenvalue of 4.10, captures 59\% of the variance in our data and is the only eigenvalue that satisfies the Kaiser criterion. The loadings in this component are strongest for social variables. Additional details on the results of PCA can be found in the SI Tables III and IV. Finally, we compute the scores for the first component for each observation and build a series of models that predict quarterly unemployment rates in provinces with and without the inclusion of this representative mobile phone variable. 

\begin{figure}[!h]
 \includegraphics[width=0.5\linewidth]{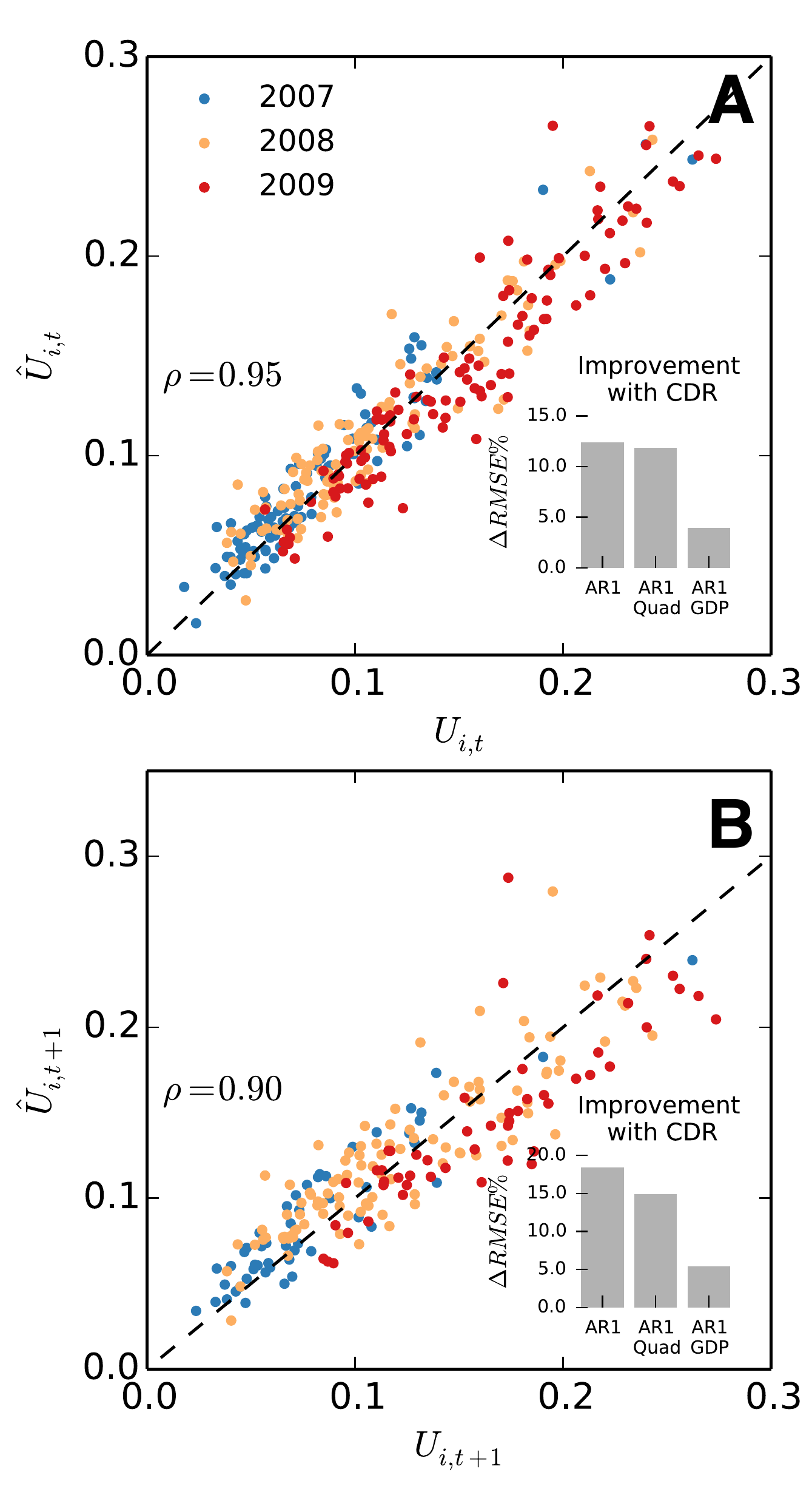}%
 \caption{\label{fig:predicting_rates}Predicting unemployment rates using mobile phone data.  We demonstrate that aggregating measurements of mobile phone behaviors associated with unemployment at the individual level also predicts unemployment rates at the province level. To make our forecasts, we train various models on data from half of the provinces and use these coefficients to predict the other half.  Panel A compares predictions of present unemployment rates to observed rates and Panel B shows predictions of unemployment one quarter ahead using an AR1 model that includes co-variates of behaviors measured using mobile phones.  Both predictions correlate strongly with actual values while changes in rates are more difficult to predict.  The insets show the percent improvement to the RMSE of predictions when mobile phone co-variates are added to various baseline model specifications. In general, the inclusion of mobile phone data reduces forecast errors by 5\% to 20\%.}
 \end{figure}

First, we \textit{predict the present} by estimating a regression of a given quarter's unemployment on calling behavior in that quarter (e.g. using phone data from Q1 to predict unemployment in Q1).  As phone data is available the day a quarter ends, this method can produce predictions weeks before survey results are tabulate and released.  Next, we \textit{predict the future} in a more traditional sense by estimating a regression on a quarter's surveyed unemployment rate using mobile phone data from last quarter as a leading indicator (e.g. phone metrics from Q1 to predict unemployment rates in Q2). This method can produce more predictions months before surveys are even conducted. See Figure 3 in the SI for a detailed timeline of data collection, release, and prediction periods. We have eight quarters of unemployment data for 52 provinces. We make and test our predictions by training our models on half of the provinces and cross-validate by testing on the other half. The groups are then switched to generate out of sample predictions for all provinces. Prediction results for an AR1 model that includes a CDR variable are plotted against actual unemployment rates in Figure \ref{fig:predicting_rates}.  We find strong correlation coefficients between predictions of predictions of present unemployment rates ($\rho=0.95$) as well as unemployment rates one quarter in the future ($\rho=0.85$).

As advocated in \cite{Lazer2014} it is important to benchmark these type of prediction algorithms against standard forecasts that use existing data. Previous work has shown that the performance of most unemployment forecasts is poor and that simple linear models  routinely outperform complicated non-linear approaches~\cite{Montgomery1998, Milas2008, Schuh2001, Stock1998} and the dynamic stochastic general equilibrium (DSGE) models aimed at simulating complex macro economic interactions~\cite{Rossi2013, Edge2010}. With this in mind, we compare predictions made with and without mobile phone covariates using three different model specifications: AR1, AR1 with a quadratic term (AR1 Quad), AR1 with a lagged national GDP covariate (AR1 GDP).  In each of these model specifications, the coefficient related to the principal component CDR score is highly significant and negative as expected given that the loadings weigh heavily on social variables that declined after a mass layoff (see tables V and VI in the SI regression results).  Moreover, adding metrics derived from mobile phone data significantly improves forecast accuracy for each model and reduces the root mean squared error of unemployment rate predictions by between 5\% and 20\% (see inserts in Figure \ref{fig:predicting_rates}). As additional checks that we are capturing true improvements, we use mobile phone data from only the first half of each quarter (before surveys are even conducted) and still achieve a 3\%-10\% improvement in forecasts. These results hold even when variants are run to include quarterly and province level fixed effects (see tables VII and VIII in the SI).

In summary, we have shown that features associated with job loss at the individual level are similarly correlated with province level changes in unemployment rates in a separate country. Moreover, we have demonstrated the ability of massive, passively collected data to identify salient features of economic shocks that can be scaled up to measure macro economic changes.  These methods allow us to predict ``present" unemployment rates two to eight weeks prior to the release of traditional estimates and predict ``future" rates up to four months ahead of official reports more accurately than using historical data alone.

\section{Discussion}
We have presented algorithms capable of identifying employment shocks at the individual, community, and societal scales from mobile phone data. These findings have great practical importance, potentially facilitating the identification of macro-economic statistics with much finer spatial granularity and faster than traditional methods of tracking the economy.  We can not only improve estimates of the current state of the economy and provide predictions faster than traditional methods, but also predict future states and correct for current uncertainties.  Moreover, with the quantity and richness of these data increasing daily, these results represent conservative estimates of its potential for predicting economic indicators.  The ability to get this information weeks to months faster than traditional methods is extremely valuable to policy and decision makers in public and private institutions. Further, it is likely that CDR data are more robust to external manipulation and less subject to service provider algorithmic changes than most social media \cite{Lazer2014}. But, just as important, the micro nature of these data allow for the development of new empirical approaches to study the effect of economic shocks on interrelated individuals.

While this study highlights the potential of new data sources to improve forecasts of critical economic indicators, we do not view these methods as a substitute for survey based approaches.  Though data quantity is increased by orders of magnitude with the collection of passively generated data from digital devices, the price of this scale is control.  The researcher no longer has the ability to precisely define which variables are collected, how they are defined, when data collection occurs making it much harder to insure data quality and integrity. In many cases, data is not collected by the researcher at all and is instead first pre-processed by the collector, introducing additional uncertainties and opportunities for contamination. Moreover, data collection itself is now conditioned on who has specific devices and services, introducing potential biases due to economic access or sorting. If policy decisions are based solely on data derived from smartphones, the segment of the population that cannot afford these devices may be underserved. 

Surveys, on the other hand, provide the researcher far more control to target specific groups, ask precise questions, and collect rich covariates.  Though the expense of creating, administering, and participating in surveys makes it difficult to collect data of the size and frequency of newer data sources, they can provide far more context about participants.  This work demonstrates the benefits of both data gathering methods and shows that hybrid models offer a way to leverage the advantages of each.  Traditional survey based forecasts are improved here, not replaced, by mobile phone data. Moving forward we hope to see more such hybrid approaches. Projects such as the Future Mobility Survey\cite{Cotrill2013} and the MIT Reality Mining project \cite{Eagle2006} bridge this gap by administering surveys via mobile devices, allowing for the collection of process generated data as well as survey based data.  These projects open the possibility to directly measure the correlation between data gathered by each approach.

The macro-economy is the complex concatenation of interdependent decisions of millions of individuals \cite{Krugman1996}. To have a measure of the activity of almost every individual in the economy, of their movements and their connections should transform our understanding of the modern economy. Moreover, the ubiquity of such data allows us to test our theories at scales large and small and all over the world with little added cost. We also note potential privacy and ethical issues regarding the inference of employment/unemployment at the individual level, with potentially dire consequences for individuals' access, for example, to financial markets. With the behavior of billions being volunteered, captured, and stored at increasingly high resolutions, these data present an opportunity to shed light on some of the biggest problems facing researchers and policy makers alike, but also represent an ethical conundrum typical of the ``big data" age.

\section{Acknowledgments}  
Jameson L. Toole received funding from the National Science Foundation Graduate Research Fellowship Program (NSF GRFP). 

\section{Author Contributions}
Jameson L. Toole designed and performed data analysis and wrote the paper.  Yu-Ru Lin designed and performed data analysis and wrote the paper. Daniel Shoag designed and  performed data analysis and wrote the paper. Erich Muehlegger designed and performed data analysis and wrote the paper. Marta C. Gonzalez provided data and edited the paper.  David Lazer designed analysis, provided data, and wrote the paper.


\begin{thebibliography}{10}
\expandafter\ifx\csname url\endcsname\relax
  \def\url#1{\texttt{#1}}\fi
\expandafter\ifx\csname urlprefix\endcsname\relax\def\urlprefix{URL }\fi
\providecommand{\bibinfo}[2]{#2}
\providecommand{\eprint}[2][]{\url{#2}}

\bibitem{Krueger2014}
\bibinfo{author}{Krueger, A.}, \bibinfo{author}{Mas, A.} \&
  \bibinfo{author}{Niu, X.}
\newblock \bibinfo{title}{The evolution of rotation group bias: Will the real
  unemployment rate please stand up?}
\newblock \bibinfo{type}{Tech. Rep.}, \bibinfo{institution}{National Bureau of
  Economic Research} (\bibinfo{year}{2014}).

\bibitem{Marcuss2007}
\bibinfo{author}{Marcuss, D.} \& \bibinfo{author}{Kane, R.~E.}
\newblock \bibinfo{title}{Us national income and product statistics born of the
  great depression and world war ii}.
\newblock \bibinfo{type}{Tech. Rep.}, \bibinfo{institution}{Bureau of Economic
  Analysis} (\bibinfo{year}{2007}).

\bibitem{Card2011}
\bibinfo{author}{Card, D.}
\newblock \bibinfo{title}{{Origins of the Unemployment Rate: The Lasting Legacy
  of Measurement without Theory}}.
\newblock \emph{\bibinfo{journal}{American Economic Review}}
  \textbf{\bibinfo{volume}{101}}, \bibinfo{pages}{552--557}
  (\bibinfo{year}{2011}).

\bibitem{Hausman2004}
\bibinfo{author}{Hausman, J.} \& \bibinfo{author}{Leibtag, E.}
\newblock \bibinfo{title}{{CPI Bias from Supercenters: Does the BLS Know that
  Wal-Mart Exists?}}
\newblock \bibinfo{type}{Tech. Rep.}, \bibinfo{institution}{National Bureau of
  Economic Research} (\bibinfo{year}{2004}).

\bibitem{Jones1999}
\bibinfo{author}{Jones, S. R.~G.} \& \bibinfo{author}{Riddell, W.~C.}
\newblock \bibinfo{title}{{The Measurement of Unemployment: An Empirical
  Approach}}.
\newblock \emph{\bibinfo{journal}{Econometrica}} \textbf{\bibinfo{volume}{67}},
  \bibinfo{pages}{147--162} (\bibinfo{year}{1999}).

\bibitem{Burda2012}
\bibinfo{author}{Burda, M.~C.}, \bibinfo{author}{Hamermesh, D.~S.} \&
  \bibinfo{author}{Stewart, J.}
\newblock \bibinfo{title}{{Cyclical Variation in Labor Hours and Productivity
  Using the ATUS}}.
\newblock \bibinfo{type}{Tech. Rep.}, \bibinfo{institution}{National Bureau of
  Economic Research} (\bibinfo{year}{2012}).

\bibitem{Tiller1994}
\bibinfo{author}{Tiller, R.} \& \bibinfo{author}{Welch, M.}
\newblock \bibinfo{title}{{Predicting the National unemployment rate that the
  "old" CPS would have produced}}.
\newblock In \emph{\bibinfo{booktitle}{Proceedings of the Section on Survey
  Research Methods, American Statistical Association}} (\bibinfo{year}{1994}).

\bibitem{Chan1999}
\bibinfo{author}{Chan, S.} \& \bibinfo{author}{Stevens, A.~H.}
\newblock \bibinfo{title}{{Employment and retirement following a late-career
  job loss}}.
\newblock \emph{\bibinfo{journal}{American Economic Review}}
  \bibinfo{pages}{211--216} (\bibinfo{year}{1999}).

\bibitem{Ruhm1999}
\bibinfo{author}{Ruhm, C.~J.}
\newblock \bibinfo{title}{{Are workers permanently scarred by job
  displacements?}}
\newblock \emph{\bibinfo{journal}{The American Economic Review}}
  \bibinfo{pages}{319--324} (\bibinfo{year}{1991}).

\bibitem{Oreopoulos2008}
\bibinfo{author}{Oreopoulos, P.}, \bibinfo{author}{Page, M.} \&
  \bibinfo{author}{Stevens, A.~H.}
\newblock \bibinfo{title}{{The intergenerational effects of worker
  displacement}}.
\newblock \emph{\bibinfo{journal}{Journal of Labor Economics}}
  \textbf{\bibinfo{volume}{26}}, \bibinfo{pages}{0--455}
  (\bibinfo{year}{2008}).

\bibitem{Sullivan2009}
\bibinfo{author}{Sullivan, D.} \& \bibinfo{author}{{Von Wachter}, T.}
\newblock \bibinfo{title}{{Job displacement and mortality: An analysis using
  administrative data}}.
\newblock \emph{\bibinfo{journal}{The Quarterly Journal of Economics}}
  \textbf{\bibinfo{volume}{124}}, \bibinfo{pages}{1265--1306}
  (\bibinfo{year}{2009}).

\bibitem{Classen2012}
\bibinfo{author}{Classen, T.~J.} \& \bibinfo{author}{Dunn, R.~A.}
\newblock \bibinfo{title}{{The effect of job loss and unemployment duration on
  suicide risk in the United States: a new look using mass-layoffs and
  unemployment duration}}.
\newblock \emph{\bibinfo{journal}{Health economics}}
  \textbf{\bibinfo{volume}{21}}, \bibinfo{pages}{338--350}
  (\bibinfo{year}{2012}).

\bibitem{Charles2001}
\bibinfo{author}{Charles, K.~K.} \& \bibinfo{author}{{Stephens Jr}, M.}
\newblock \bibinfo{title}{{Job displacement, disability, and divorce}}.
\newblock \bibinfo{type}{Tech. Rep.}, \bibinfo{institution}{National bureau of
  economic research} (\bibinfo{year}{2001}).

\bibitem{Krueger2011}
\bibinfo{author}{Krueger, A.~B.}, \bibinfo{author}{Mueller, A.},
  \bibinfo{author}{Davis, S.~J.} \& \bibinfo{author}{Sahin, A.}
\newblock \bibinfo{title}{Job search, emotional well-being, and job finding in
  a period of mass unemployment: Evidence from high frequency longitudinal data
  [with comments and discussion]}.
\newblock \emph{\bibinfo{journal}{Brookings Papers on Economic Activity}}
  \bibinfo{pages}{pp. 1--81} (\bibinfo{year}{2011}).

\bibitem{Choi2009}
\bibinfo{author}{Choi, H.} \& \bibinfo{author}{Varian, H.}
\newblock \bibinfo{title}{{Predicting initial claims for unemployment
  benefits}}.
\newblock \emph{\bibinfo{journal}{Google Inc}}  (\bibinfo{year}{2009}).

\bibitem{Vespignani2009}
\bibinfo{author}{Vespignani, A.}
\newblock \bibinfo{title}{{Predicting the behavior of techno-social systems.}}
\newblock \emph{\bibinfo{journal}{Science}} \textbf{\bibinfo{volume}{325}},
  \bibinfo{pages}{425--8} (\bibinfo{year}{2009}).

\bibitem{Lazer2009}
\bibinfo{author}{Lazer, D.} \emph{et~al.}
\newblock \bibinfo{title}{{Computational Social Science}}.
\newblock \emph{\bibinfo{journal}{Science}} \textbf{\bibinfo{volume}{323}},
  \bibinfo{pages}{721--723} (\bibinfo{year}{2009}).

\bibitem{Henderson2012}
\bibinfo{author}{Henderson, J.~V.}, \bibinfo{author}{Storeygard, A.} \&
  \bibinfo{author}{Weil, D.~N.}
\newblock \bibinfo{title}{{Measuring Economic Growth from Outerspace}}.
\newblock \emph{\bibinfo{journal}{The American Economic Review}}
  \textbf{\bibinfo{volume}{102}}, \bibinfo{pages}{994--1028}
  (\bibinfo{year}{2012}).

\bibitem{Hayo2005}
\bibinfo{author}{Hayo, B.} \& \bibinfo{author}{Kutan, A.~M.}
\newblock \bibinfo{title}{{The impact of news, oil prices, and global market
  developments on Russian financial markets1}}.
\newblock \emph{\bibinfo{journal}{Economics of Transition}}
  \textbf{\bibinfo{volume}{13}}, \bibinfo{pages}{373--393}
  (\bibinfo{year}{2005}).

\bibitem{Lavrenko2000}
\bibinfo{author}{Lavrenko, V.} \emph{et~al.}
\newblock \bibinfo{title}{{Mining of concurrent text and time series}}.
\newblock In \emph{\bibinfo{booktitle}{KDD-2000 Workshop on Text Mining}},
  \bibinfo{pages}{37--44} (\bibinfo{organization}{Citeseer},
  \bibinfo{year}{2000}).

\bibitem{Schumaker2009}
\bibinfo{author}{Schumaker, R.~P.} \& \bibinfo{author}{Chen, H.}
\newblock \bibinfo{title}{{Textual analysis of stock market prediction using
  breaking financial news: The AZFin text system}}.
\newblock \emph{\bibinfo{journal}{ACM Transactions on Information Systems
  (TOIS)}} \textbf{\bibinfo{volume}{27}}, \bibinfo{pages}{12}
  (\bibinfo{year}{2009}).

\bibitem{Ettredge2005}
\bibinfo{author}{Ettredge, M.}, \bibinfo{author}{Gerdes, J.} \&
  \bibinfo{author}{Karuga, G.}
\newblock \bibinfo{title}{{Using web-based search data to predict macroeconomic
  statistics}}.
\newblock \emph{\bibinfo{journal}{Communications of the ACM}}
  \textbf{\bibinfo{volume}{48}}, \bibinfo{pages}{87--92}
  (\bibinfo{year}{2005}).

\bibitem{Askitas2009}
\bibinfo{author}{Askitas, N.} \& \bibinfo{author}{Zimmermann, K.~F.}
\newblock \bibinfo{title}{{Google econometrics and unemployment forecasting}}
  (\bibinfo{year}{2009}).

\bibitem{Eagle2006}
\bibinfo{author}{Eagle, N.} \& \bibinfo{author}{Pentland, A.}
\newblock \bibinfo{title}{{Reality mining: Sensing complex social systems}}.
\newblock \emph{\bibinfo{journal}{Personal and Ubiquitous Computing}}
  \textbf{\bibinfo{volume}{10}}, \bibinfo{pages}{255--268}
  (\bibinfo{year}{2006}).

\bibitem{Suhoy2009}
\bibinfo{author}{Suhoy, T.}
\newblock \emph{\bibinfo{title}{{Query indices and a 2008 downturn: Israeli
  data}}} (\bibinfo{publisher}{Research Department, Bank of Israel},
  \bibinfo{year}{2009}).

\bibitem{Antenucci2014}
\bibinfo{author}{Antenucci, D.}, \bibinfo{author}{Cafarella, M.},
  \bibinfo{author}{Levenstein, M.}, \bibinfo{author}{R\'{e}, C.} \&
  \bibinfo{author}{Shapiro, M.~D.}
\newblock \bibinfo{title}{{Using Social Media to Measure Labor Market Flows}}.
\newblock \bibinfo{type}{Tech. Rep.}, \bibinfo{institution}{National Bureau of
  Economic Research} (\bibinfo{year}{2014}).

\bibitem{Goel2010}
\bibinfo{author}{Goel, S.}, \bibinfo{author}{Hofman, J.~M.},
  \bibinfo{author}{Lahaie, S.}, \bibinfo{author}{Pennock, D.~M.} \&
  \bibinfo{author}{Watts, D.~J.}
\newblock \bibinfo{title}{{Predicting consumer behavior with Web search}}.
\newblock \emph{\bibinfo{journal}{Proceedings of the National Academy of
  Sciences}} \textbf{\bibinfo{volume}{107}}, \bibinfo{pages}{17486--17490}
  (\bibinfo{year}{2010}).

\bibitem{Gelman2014}
\bibinfo{author}{Gelman, M.}, \bibinfo{author}{Kariv, S.},
  \bibinfo{author}{Shapiro, M.~D.}, \bibinfo{author}{Silverman, D.} \&
  \bibinfo{author}{Tadelis, S.}
\newblock \bibinfo{title}{{Harnessing naturally occurring data to measure the
  response of spending to income}}.
\newblock \emph{\bibinfo{journal}{Science}} \textbf{\bibinfo{volume}{345}},
  \bibinfo{pages}{212--215} (\bibinfo{year}{2014}).

\bibitem{Choi2012}
\bibinfo{author}{Choi, H.} \& \bibinfo{author}{Varian, H.}
\newblock \bibinfo{title}{{Predicting the Present with Google Trends}}.
\newblock \emph{\bibinfo{journal}{Economic Record}}
  \textbf{\bibinfo{volume}{88}}, \bibinfo{pages}{2--9} (\bibinfo{year}{2012}).

\bibitem{Ginsberg2008}
\bibinfo{author}{Ginsberg, J.} \emph{et~al.}
\newblock \bibinfo{title}{{Detecting influenza epidemics using search engine
  query data}}.
\newblock \emph{\bibinfo{journal}{Nature}} \textbf{\bibinfo{volume}{457}},
  \bibinfo{pages}{1012--1014} (\bibinfo{year}{2008}).

\bibitem{Aramaki2011}
\bibinfo{author}{Aramaki, E.}, \bibinfo{author}{Maskawa, S.} \&
  \bibinfo{author}{Morita, M.}
\newblock \bibinfo{title}{{Twitter catches the flu: detecting influenza
  epidemics using Twitter}}.
\newblock In \emph{\bibinfo{booktitle}{Proceedings of the Conference on
  Empirical Methods in Natural Language Processing}},
  \bibinfo{pages}{1568--1576} (\bibinfo{organization}{Association for
  Computational Linguistics}, \bibinfo{year}{2011}).

\bibitem{Chew2010}
\bibinfo{author}{Chew, C.} \& \bibinfo{author}{Eysenbach, G.}
\newblock \bibinfo{title}{{Pandemics in the age of Twitter: content analysis of
  Tweets during the 2009 H1N1 outbreak}}.
\newblock \emph{\bibinfo{journal}{PloS one}} \textbf{\bibinfo{volume}{5}},
  \bibinfo{pages}{e14118} (\bibinfo{year}{2010}).

\bibitem{Culotta2010}
\bibinfo{author}{Culotta, A.}
\newblock \bibinfo{title}{{Towards detecting influenza epidemics by analyzing
  Twitter messages}}.
\newblock In \emph{\bibinfo{booktitle}{Proceedings of the first workshop on
  social media analytics}}, \bibinfo{pages}{115--122}
  (\bibinfo{organization}{ACM}, \bibinfo{year}{2010}).

\bibitem{Gomide2011}
\bibinfo{author}{Gomide, J.} \emph{et~al.}
\newblock \bibinfo{title}{{Dengue surveillance based on a computational model
  of spatio-temporal locality of Twitter}}.
\newblock In \emph{\bibinfo{booktitle}{Proceedings of the 3rd International Web
  Science Conference}}, \bibinfo{pages}{3} (\bibinfo{organization}{ACM},
  \bibinfo{year}{2011}).

\bibitem{Quincey2010}
\bibinfo{author}{de~Quincey, E.} \& \bibinfo{author}{Kostkova, P.}
\newblock \bibinfo{title}{{Early warning and outbreak detection using social
  networking websites: The potential of twitter}}.
\newblock In \emph{\bibinfo{booktitle}{Electronic healthcare}},
  \bibinfo{pages}{21--24} (\bibinfo{publisher}{Springer},
  \bibinfo{year}{2010}).

\bibitem{Zamite2011}
\bibinfo{author}{Zamite, J.~a.}, \bibinfo{author}{Silva, F. A.~B.},
  \bibinfo{author}{Couto, F.} \& \bibinfo{author}{Silva, M.~J.}
\newblock \bibinfo{title}{{MEDCollector: Multisource epidemic data collector}}.
\newblock In \emph{\bibinfo{booktitle}{Transactions on large-scale data-and
  knowledge-centered systems IV}}, \bibinfo{pages}{40--72}
  (\bibinfo{publisher}{Springer}, \bibinfo{year}{2011}).

\bibitem{Signorini2011}
\bibinfo{author}{Signorini, A.}, \bibinfo{author}{Segre, A.~M.} \&
  \bibinfo{author}{Polgreen, P.~M.}
\newblock \bibinfo{title}{{The use of Twitter to track levels of disease
  activity and public concern in the US during the influenza A H1N1 pandemic}}.
\newblock \emph{\bibinfo{journal}{PloS one}} \textbf{\bibinfo{volume}{6}},
  \bibinfo{pages}{e19467} (\bibinfo{year}{2011}).

\bibitem{Lazer2014}
\bibinfo{author}{Lazer, D.}, \bibinfo{author}{Kennedy, R.},
  \bibinfo{author}{King, G.} \& \bibinfo{author}{Vespignani, A.}
\newblock \bibinfo{title}{{Big data. The parable of Google Flu: traps in big
  data analysis.}}
\newblock \emph{\bibinfo{journal}{Science (New York, N.Y.)}}
  \textbf{\bibinfo{volume}{343}}, \bibinfo{pages}{1203--5}
  (\bibinfo{year}{2014}).

\bibitem{Sakaki2010}
\bibinfo{author}{Sakaki, T.}, \bibinfo{author}{Okazaki, M.} \&
  \bibinfo{author}{Matsuo, Y.}
\newblock \bibinfo{title}{Earthquake shakes twitter users: real-time event
  detection by social sensors}.
\newblock In \emph{\bibinfo{booktitle}{Proceedings of the 19th international
  conference on World wide web}}, \bibinfo{pages}{851--860}
  (\bibinfo{organization}{ACM}, \bibinfo{year}{2010}).

\bibitem{Guy2010}
\bibinfo{author}{Guy, M.}, \bibinfo{author}{Earle, P.},
  \bibinfo{author}{Ostrum, C.}, \bibinfo{author}{Gruchalla, K.} \&
  \bibinfo{author}{Horvath, S.}
\newblock \bibinfo{title}{{Integration and dissemination of citizen reported
  and seismically derived earthquake information via social network
  technologies}}.
\newblock In \emph{\bibinfo{booktitle}{Advances in intelligent data analysis
  IX}}, \bibinfo{pages}{42--53} (\bibinfo{publisher}{Springer},
  \bibinfo{year}{2010}).

\bibitem{Earle2012}
\bibinfo{author}{Earle, P.}, \bibinfo{author}{Bowden, D.} \&
  \bibinfo{author}{Guy, M.}
\newblock \bibinfo{title}{Twitter earthquake detection: earthquake monitoring
  in a social world}.
\newblock \emph{\bibinfo{journal}{Annals of Geophysics}}
  \textbf{\bibinfo{volume}{54}}, \bibinfo{pages}{708--715}
  (\bibinfo{year}{2012}).

\bibitem{Barabasi2005}
\bibinfo{author}{Barab\'{a}si, A.-L.}
\newblock \bibinfo{title}{{The origin of bursts and heavy tails in human
  dynamics}}.
\newblock \emph{\bibinfo{journal}{Nature}} \textbf{\bibinfo{volume}{435}},
  \bibinfo{pages}{207--211} (\bibinfo{year}{2005}).

\bibitem{Rybski2009}
\bibinfo{author}{Rybski, D.}, \bibinfo{author}{Buldyrev, S.~V.},
  \bibinfo{author}{Havlin, S.}, \bibinfo{author}{Liljeros, F.} \&
  \bibinfo{author}{Makse, H.~A.}
\newblock \bibinfo{title}{{Scaling laws of human interaction activity}}.
\newblock \emph{\bibinfo{journal}{Proceedings of the National Academy of
  Sciences}} \textbf{\bibinfo{volume}{106}}, \bibinfo{pages}{12640--12645}
  (\bibinfo{year}{2009}).

\bibitem{Onnela2011}
\bibinfo{author}{Onnela, J.-P.}, \bibinfo{author}{Arbesman, S.},
  \bibinfo{author}{Gonz\'{a}lez, M.~C.}, \bibinfo{author}{Barab\'{a}si, A.-L.}
  \& \bibinfo{author}{Christakis, N.~A.}
\newblock \bibinfo{title}{{Geographic Constraints on Social Network Groups}}.
\newblock \emph{\bibinfo{journal}{PLoS ONE}} \textbf{\bibinfo{volume}{6}},
  \bibinfo{pages}{7} (\bibinfo{year}{2011}).

\bibitem{Onnela2007}
\bibinfo{author}{Onnela, J.-P.} \emph{et~al.}
\newblock \bibinfo{title}{{Structure and tie strengths in mobile communication
  networks.}}
\newblock \emph{\bibinfo{journal}{Proceedings of the National Academy of
  Sciences of the United States of America}} \textbf{\bibinfo{volume}{104}},
  \bibinfo{pages}{7332--7336} (\bibinfo{year}{2007}).

\bibitem{Cho2011}
\bibinfo{author}{Cho, E.}, \bibinfo{author}{Myers, S.~A.} \&
  \bibinfo{author}{Leskovec, J.}
\newblock \bibinfo{title}{{Friendship and mobility}}.
\newblock In \emph{\bibinfo{booktitle}{Proceedings of the 17th ACM SIGKDD
  international conference on Knowledge discovery and data mining KDD 11}}, KDD
  '11, \bibinfo{pages}{1082} (\bibinfo{publisher}{ACM Press},
  \bibinfo{year}{2011}).

\bibitem{Goncalves2008}
\bibinfo{author}{Goncalves, B.} \& \bibinfo{author}{Ramasco, J.~J.}
\newblock \bibinfo{title}{{Human dynamics revealed through Web analytics}}.
\newblock \emph{\bibinfo{journal}{Physical Review E}}
  \textbf{\bibinfo{volume}{78}}, \bibinfo{pages}{7} (\bibinfo{year}{2008}).

\bibitem{Bond2012}
\bibinfo{author}{Bond, R.~M.} \emph{et~al.}
\newblock \bibinfo{title}{A 61-million-person experiment in social influence
  and political mobilization}.
\newblock \emph{\bibinfo{journal}{Nature}} \textbf{\bibinfo{volume}{489}},
  \bibinfo{pages}{295--298} (\bibinfo{year}{2012}).

\bibitem{Balcan2009}
\bibinfo{author}{Balcan, D.} \emph{et~al.}
\newblock \bibinfo{title}{{Multiscale mobility networks and the spatial
  spreading of infectious diseases.}}
\newblock \emph{\bibinfo{journal}{Proceedings of the National Academy of
  Sciences of the United States of America}} \textbf{\bibinfo{volume}{106}},
  \bibinfo{pages}{21484--9} (\bibinfo{year}{2009}).

\bibitem{Han2011}
\bibinfo{author}{Han, X.-P.}, \bibinfo{author}{Hao, Q.}, \bibinfo{author}{Wang,
  B.-H.} \& \bibinfo{author}{Zhou, T.}
\newblock \bibinfo{title}{{Origin of the scaling law in human mobility:
  Hierarchy of traffic systems}}.
\newblock \emph{\bibinfo{journal}{Physical Review E}}
  \textbf{\bibinfo{volume}{83}}, \bibinfo{pages}{2--6} (\bibinfo{year}{2011}).

\bibitem{Brockmann2006}
\bibinfo{author}{Brockmann, D.}, \bibinfo{author}{Hufnagel, L.} \&
  \bibinfo{author}{Geisel, T.}
\newblock \bibinfo{title}{{The scaling laws of human travel}}.
\newblock \emph{\bibinfo{journal}{Nature}} \textbf{\bibinfo{volume}{439}},
  \bibinfo{pages}{462--5} (\bibinfo{year}{2006}).

\bibitem{Gonzalez2008}
\bibinfo{author}{Gonz\'{a}lez, M.~C.}, \bibinfo{author}{Hidalgo, C.~A.} \&
  \bibinfo{author}{Barab\'{a}si, A.-L.}
\newblock \bibinfo{title}{{Understanding individual human mobility patterns.}}
\newblock \emph{\bibinfo{journal}{Nature}} \textbf{\bibinfo{volume}{453}},
  \bibinfo{pages}{779--782} (\bibinfo{year}{2008}).

\bibitem{Song2010a}
\bibinfo{author}{Song, C.}, \bibinfo{author}{Qu, Z.}, \bibinfo{author}{Blumm,
  N.} \& \bibinfo{author}{Barab\'{a}si, A.-L.}
\newblock \bibinfo{title}{{Limits of predictability in human mobility.}}
\newblock \emph{\bibinfo{journal}{Science}} \textbf{\bibinfo{volume}{327}},
  \bibinfo{pages}{1018--1021} (\bibinfo{year}{2010}).

\bibitem{Song2010b}
\bibinfo{author}{Song, C.}, \bibinfo{author}{Koren, T.}, \bibinfo{author}{Wang,
  P.} \& \bibinfo{author}{Barab\'{a}si, A.-L.}
\newblock \bibinfo{title}{{Modelling the scaling properties of human
  mobility}}.
\newblock \emph{\bibinfo{journal}{Nature Physics}}
  \textbf{\bibinfo{volume}{6}}, \bibinfo{pages}{818--823}
  (\bibinfo{year}{2010}).

\bibitem{Deville2014}
\bibinfo{author}{Deville, P.} \emph{et~al.}
\newblock \bibinfo{title}{{Dynamic population mapping using mobile phone
  data}}.
\newblock \emph{\bibinfo{journal}{Proceedings of the National Academy of
  Sciences}} \bibinfo{pages}{1408439111--} (\bibinfo{year}{2014}).
  
 \bibitem{Yan2014}
\bibinfo{author}{Yan, Xiao-Yong}, \bibinfo{author}{Zhao, Chen}, \bibinfo{author}{Fan, Ying}, \bibinfo{author}{Di, Zengru}, \& \bibinfo{author} {Wang, Wen-Xu}
\newblock \bibinfo{title}{Universal predictability of mobility patterns in cities}
\newblock \emph{\bibinfo{journal}{Journal of the Royal Society Interface}}
\textbf{\bibinfo{volume}{11}}, (\bibinfo{year}{2014}).

\bibitem{Bagrow2011}
\bibinfo{author}{Bagrow, J.~P.}, \bibinfo{author}{Wang, D.} \&
  \bibinfo{author}{Barab\'{a}si, A.-L.}
\newblock \bibinfo{title}{{Collective Response of Human Populations to
  Large-Scale Emergencies}}.
\newblock \emph{\bibinfo{journal}{PLoS ONE}} \textbf{\bibinfo{volume}{6}},
  \bibinfo{pages}{8} (\bibinfo{year}{2011}).

\bibitem{Blumenstock2011}
\bibinfo{author}{Blumenstock, J.~E.}, \bibinfo{author}{Fafchamps, M.} \&
  \bibinfo{author}{Eagle, N.}
\newblock \bibinfo{title}{{Risk and Reciprocity Over the Mobile Phone Network:
  Evidence from Rwanda}}.
\newblock \emph{\bibinfo{journal}{SSRN Electronic Journal}}
  (\bibinfo{year}{2011}).

\bibitem{Lu2012}
\bibinfo{author}{Lu, X.}, \bibinfo{author}{Bengtsson, L.} \&
  \bibinfo{author}{Holme, P.}
\newblock \bibinfo{title}{{Predictability of population displacement after the
  2010 Haiti earthquake.}}
\newblock \emph{\bibinfo{journal}{Proceedings of the National Academy of
  Sciences of the United States of America}} \textbf{\bibinfo{volume}{109}},
  \bibinfo{pages}{11576--81} (\bibinfo{year}{2012}).

\bibitem{Calabrese2011}
\bibinfo{author}{Calabrese, F.}, \bibinfo{author}{Colonna, M.},
  \bibinfo{author}{Lovisolo, P.}, \bibinfo{author}{Parata, D.} \&
  \bibinfo{author}{Ratti, C.}
\newblock \bibinfo{title}{{Real-Time Urban Monitoring Using Cell Phones: A Case
  Study in Rome}} \textbf{\bibinfo{volume}{12}}, \bibinfo{pages}{141--151}
  (\bibinfo{year}{2011}).

\bibitem{Calabrese2006}
\bibinfo{author}{Calabrese, F.} \& \bibinfo{author}{Ratti, C.}
\newblock \bibinfo{title}{{Real Time Rome}}.
\newblock \emph{\bibinfo{journal}{NETCOM}} \textbf{\bibinfo{volume}{20}},
  \bibinfo{pages}{247--258} (\bibinfo{year}{2006}).

\bibitem{Montgomery1998}
\bibinfo{author}{Montgomery, A.~L.}, \bibinfo{author}{Zarnowitz, V.},
  \bibinfo{author}{Tsay, R.~S.} \& \bibinfo{author}{Tiao, G.~C.}
\newblock \bibinfo{title}{Forecasting the us unemployment rate}.
\newblock \emph{\bibinfo{journal}{Journal of the American Statistical
  Association}} \textbf{\bibinfo{volume}{93}}, \bibinfo{pages}{478--493}
  (\bibinfo{year}{1998}).

\bibitem{Milas2008}
\bibinfo{author}{Milas, C.} \& \bibinfo{author}{Rothman, P.}
\newblock \bibinfo{title}{{Out-of-sample forecasting of unemployment rates with
  pooled STVECM forecasts}}.
\newblock \emph{\bibinfo{journal}{International Journal of Forecasting}}
  \textbf{\bibinfo{volume}{24}}, \bibinfo{pages}{101--121}
  (\bibinfo{year}{2008}).

\bibitem{Schuh2001}
\bibinfo{author}{Schuh, S.}
\newblock \bibinfo{title}{{An evaluation of recent macroeconomic forecast
  errors}}.
\newblock \emph{\bibinfo{journal}{New England Economic Review}}
  \bibinfo{pages}{35--56} (\bibinfo{year}{2001}).

\bibitem{Stock1998}
\bibinfo{author}{Stock, J.~H.} \& \bibinfo{author}{Watson, M.~W.}
\newblock \bibinfo{title}{{A Comparison of Linear and Nonlinear Univariate
  Models for Forecasting Macroeconomic Time Series}}.
\newblock \bibinfo{type}{Tech. Rep.}, \bibinfo{institution}{National Bureau of
  Economic Research} (\bibinfo{year}{1998}).

\bibitem{Rossi2013}
\bibinfo{author}{Rossi, B.}
\newblock \bibinfo{title}{{Do DSGE Models Forecast More Accurately
  Out-of-Sample than VAR Models?}}
\newblock \emph{\bibinfo{journal}{Advances in Econometrics}}
  \textbf{\bibinfo{volume}{32}}, \bibinfo{pages}{27--79}
  (\bibinfo{year}{2013}).

\bibitem{Edge2010}
\bibinfo{author}{Edge, R.~M.}, \bibinfo{author}{G{\"u}rkaynak, R.~S.},
  \bibinfo{author}{REIS, R.} \& \bibinfo{author}{SIMS, C.~A.}
\newblock \bibinfo{title}{How useful are estimated dsge model forecasts for
  central bankers?[with comments and discussion]}.
\newblock \emph{\bibinfo{journal}{Brookings Papers on Economic Activity}}
  \bibinfo{pages}{209--259} (\bibinfo{year}{2010}).
  
\bibitem{Cotrill2013}
\bibinfo{author}{Cottrill, C. D.}, \bibinfo{author}{Pereira, F. C.}, \bibinfo{author}{Zhao, F., Dias}, \bibinfo{author}{I. F., Lim}, \bibinfo{author}{H. B., Ben-Akiva, M. E.}, \& \bibinfo{author}{Zegras, P. C.}
\newblock \bibinfo{title}{Future mobility survey.} 
\newblock \emph{\bibinfo{journal}{Transportation Research Record: Journal of the Transportation Research Board}}
\textbf{\bibinfo{volume}{2354}}, \bibinfo{pages}{59-67} (\bibinfo{year}{2013}).

\bibitem{Krugman1996}
\bibinfo{author}{Krugman, P.~R.}
\newblock \emph{\bibinfo{title}{{The Self Organizing Economy}}}
  (\bibinfo{publisher}{Blackwell Publishers}, \bibinfo{year}{1996}).

\end{thebibliography}

\section{Supplementary Information}
\subsection{Materials and Methods}
\begin{description}
\item[CDR Data set 1 (D1)] We analyze call detail records (CDRs) from two industrialized European countries.  In the first country, we obtain data on 1.95 million users from a service provider with roughly 15\% market share.  The data run for 15 months across the years 2006 and 2007, with the exception of a gap between August and September 2006.  Each call record includes a de-identified caller and recipient IDs, the locations of the caller's and recipient's cell towers and the length of the call. Caller or recipients on other network carriers are assigned random IDs. There are approximately 1.95 million individuals identified in the data, 453 million calls, and 16 million hours of call time. The median user makes or receives 90 calls per months.

\item[CDR Data set 2 (D2)] The second data set contains 10 million users (roughly 20\% market share) within a single country over three years of activity.  Like D1, each billing record for voice and text services, contains the unique identifiers of the caller placing the call and the callee receiving the call, an identifier for the cellular antenna (tower) that handled the call, and the date and time when the call was placed. Coupled with a data set describing the locations (latitude and longitude) of cellular towers, we have the approximate location of the caller when placing the call. For this work we do not distinguish between voice calls and text messages, and refer to either communication type as a ``call." However, we also possess identification numbers for phones that are outside the service provider but that make or receive calls to users within the company. While we do not possess any other information about these lines, nor anything about their users or calls that are made to other numbers outside the service provider, we do have records pertaining to all calls placed to or from those ID numbers involving subscribers covered by our data set. Thus egocentric networks between users within the company and their immediate neighbors only are complete. This information was used to generate egocentric communication networks and to compute the features described in the main text. From this data set, we generate a random sample population of $k$ users for each of the provinces, and track each user's call history during our 27-month tracking period (from December 2006 to March 2009). We discuss how the sample size is chosen in a following subsection.  Finally, we note that due to an error in data extraction from the provider, we are missing data for Q4 in 2007.
\end{description}

The use of CDR data to study mobility and social behaviors on a massive scale is becoming increasingly common. In addition to its large scale, its format is generally consistent between countries and mobile operators.  In the context of this study, each mobile phone data set contains five columns: 1) an anonymized, unique identifier for the caller, 2) the ID of the tower through which the caller;s call was routed, 3) an anonymized, unique identifier of the receiver of the call, 4) the ID of the tower through which the receiver's call was routed, and 5) the timestamp down to the second which the call was initiated.  In order to obtain the location of both caller and receiver, data is restricted to only calls between members of the same mobile operator. The tower IDs reflect the tower used upon starting the call and we have no information on changes in location that may be made during the call. We also obtain a list of latitudes and longitudes marking the coordinates of each tower.  Although calls are generally believed to be routed through the tower that is geographically closest to the phone, this may not be the case if the signal is obscured by buildings or topology.  For this reason, we consider a cluster of towers near the geographic area in question instead of a single tower.

To ensure privacy of mobile phone subscribers, all identifiers were anonymized before we received the data and no billing or demographic information was provided on individuals or in aggregate.

\subsection{Filtering CDR Data}
We limit our sample to mobile phone users who either make or receive at least ten calls connecting through one of the three cell towers closest to the manufacturing plant of interest.  In addition, we require that users make at least one call in each month spanned by a given data set to ensure users are still active.

\subsection{Manufacturing plant closure}
We gather information on a large manufacturing plant closing that affected a small community within the service provider's territory from news articles and administrative sources collected by the country's labor statistics bureau.  The plant closure occurred in December 2006 and involved 1,100 employees at an auto-parts manufacturing plant in a town of roughly 15,000 people. 

\subsection{Town Level Structural Break Model}
We model the pre-closure daily population of the town as consisting of three segments: a fraction of non-resident plant workers $\gamma$, a fraction of resident workers $\mu$, and a fraction of non-workers normalized to $(1-\gamma- \mu)$. We postulate that each individual $i$ has a flow probability of making or receiving a call at every moment $p_i$. Workers spend a fraction $\psi$ of their day at their jobs and thus make, in expectation, $p_i \psi$ call on a given day during work hours. When losing their job in the town, both resident and non-resident workers are re-matched in national, not local, labor market.

Given this model, the expected daily number of cell phone subscribers making or receiving calls from the three towers serving the plant and town:

\[
	vol =
		\begin{cases}
			\gamma\psi\bar{p} + (1-\gamma)\bar{p} & \text{for } t < t_{layoff} \\
			\mu(1-\psi)\bar{p} + (1-\gamma-\mu)\bar{p} & \text{for } t \geq t_{layoff}
		\end{cases}
\]

This model predicts a discrete break in daily volume from the towers proximate to the plant of  $(\gamma + \mu)\psi\bar{p}$ at the date $t_{layoff}$.  For workers, the predicted percentage change in call volume from these towers is $\frac{(\gamma + \mu)\psi \bar{p}}{(\mu\bar{p}+\gamma\psi\bar{p}}$. Non-workers experience no change.
\pagebreak
\subsection{Individual Structural Break Model}
\begin{figure}[ht]
 \includegraphics[width=0.6\linewidth]{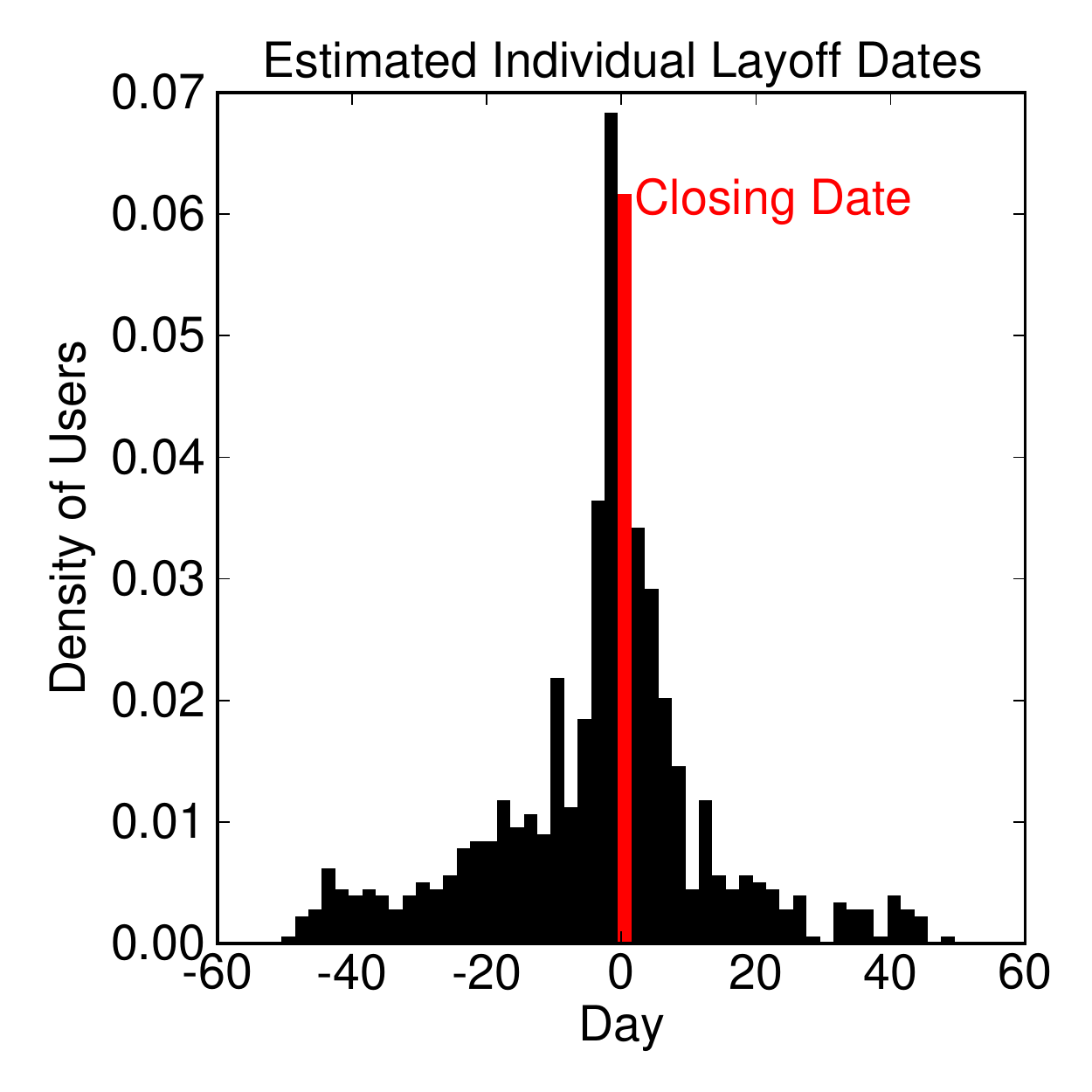}%
 \caption{\label{fig:structural_breaks}We plot the distribution of break dates for the structural break model estimated for individuals.  We find a strong, statistically significant peak centered on the reported closure date (red) with far fewer breaks on other, placebo dates.  This is consistent with both our community wide model as well as the Bayesian model presented above.}
 \end{figure}

We fit a model similar to the community structural break model to data for each individual user, $i$, based on the probability they made a call from the town on each day. For each individual, we use the non-linear estimator to select a break date $t_{layoff}^i$ , and constant pre- and post- break daily probabilities $p_{i,t<t_{layoff}^i}$ and $p_{i,t>t_{layoff}^i}$   to minimize the squared deviation from each individuals' data. Figure \ref{fig:structural_breaks} plots the distribution of break-dates for individuals. As expected, there is a statistically significant spike in the number of individuals experiencing a break in the probability of making a call from the town at the time of the closure and significantly fewer breaks on other, placebo dates. These two methods provide independent, yet complementary ways of detecting mass layoffs in mobile phone data.

\subsection{Bayesian Estimation}
\begin{figure}[ht]
 \includegraphics[width=0.5\linewidth]{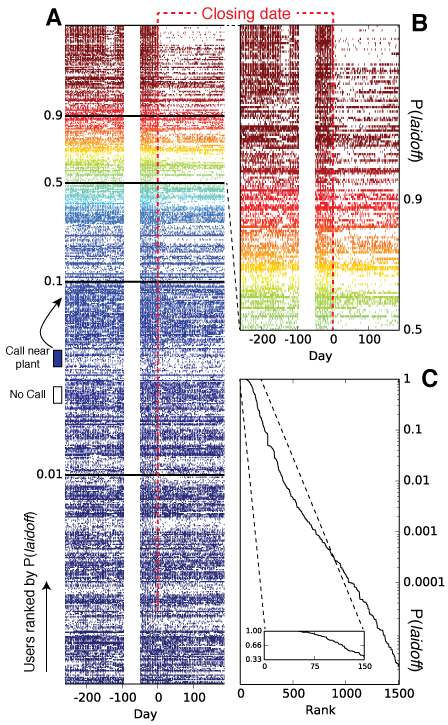}%
 \caption{\label{fig:affected_individuals}Identifying affected individuals. A) Each user is represented by a row where we fill in a day as colored if a call was made near the plant on that day.  White space marks the absence of calls. Rows are sorted by the assigned probability of that user being laid off.  B) A closer view of the users identified as mostly to have been laid off reveals a sharp cut off in days on which calls were made from the plant.  C) An inverse cumulative distribution of assigned probability weights. The insert shows an enlarged view at the probability distribution for the 150 individuals deemed most likely to have been laid off.}
 \end{figure}

On an individual level, Figure \ref{fig:affected_individuals}A shows days on which each user makes a call near the plant ranked from highest to lowest probability weight. Figure \ref{fig:affected_individuals}B provides greater detail for users probability weights between 50\% and 100\%.  Users highly suspected of being laid off demonstrate a sharp decline in the number of days they make calls near the plant following the reported closure date.  Figure \ref{fig:affected_individuals}C graphs the inverse cumulative distribution of probability weights.  While we do not have ground-truth evidence that any of these mobile phone users was laid off, we find more support for our hypothesis by examining a two week period roughly 125 days prior to the plant closure.

Figures \ref{fig:affected_individuals}A and \ref{fig:affected_individuals}B illustrate that the call patterns of users assigned the highest probabilities change significantly after the plant closure.  These users make calls from the town on a consistent basis before the layoff, but make significantly fewer calls from the town afterwards. In contrast, the call patterns of users assigned the lowest weights do not change following the plant closure. In aggregate, we assign 143 users probability weights between 50\% and 100\%.  This represents 13\% of the pre-closure plant workforce €" this fraction compares closely with the roughly 15\% national market share of the service provider.  

\subsection{The European Labor Force Survey}
Each quarter, many European countries are required to conduct labor force surveys to measure important economic indicators like the unemployment rates studied here.  In person or telephone interviews concerning employment status are conducted on a sample size of less than 0.5\% of the population.  Moreover, participants are only asked to provide responses about their employment status during a 1 week period in the quarter.

These ``microdata" surveys are then aggregated at the province and national levels.  Confirmed labor force reports and statistics for a particular quarter are released roughly 14 weeks after the quarter has ended.  For example, Q1 of 2012 begins January 1st, 2012 and ends March 31st, 2012.  The survey data is analyzed and unemployment numbers are released between two and three weeks following the end of the quarter.  These numbers, however, are unconfirmed and subject to revisions which can occur at any time in the following quarters.

\begin{figure*}[ht]
 \includegraphics[width=1\linewidth]{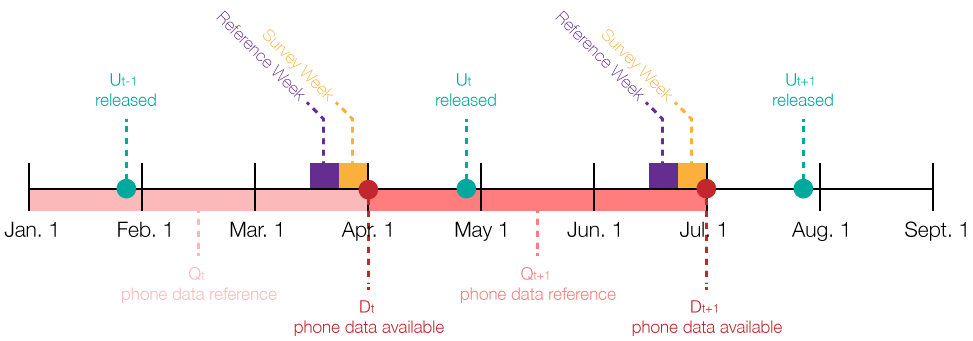}%
 \caption{\label{fig:timeline}A timeline showing the various data collection and reporting periods.  Traditional survey method perform surveys over the course of a single week per quarter, asking participants about their employment status during a single reference week.  Unofficial survey results, subject to revision are then released a few weeks following the end of the quarter.  Mobile phone data, however, is continually collected throughout the quarter and is available for analysis at any time during the period.  Analysis of a given quarter can be performed and made available immediately following the end of the month.}
 \end{figure*}

\subsection{The Effect of Job Loss on Call Volumes}
We measure the effect of job loss on six properties of an individual's social behavior and three mobility metrics.

\subsubsection{CDR Metrics}
\begin{description}
\item[Calls:] The total number of calls made and received by a user in a given month.
\item[Incoming:]The number of calls received by a user in a given month.
\item[Outgoing:] The number of calls made by a user in a given month
\item[Contacts:] The number of unique individuals contacted by a user each month.  Includes calls made and received.
\item[To Town:] The fraction of a user's calls made each month to another user who is physically located in the town of the plant closure at the time the call was made.
\item[Churn:] The fraction of a user's contacts called in the previous month that was not called in the current month.  Let $C_t$ be the set of users called in month $t$.  Churn is then calculated as: $churn_t = 1-\frac{|C_{t-1}-C_t|}{|C_t-1|}$.
\item[Towers:] The number of unique towers visited by a user each month.
\item[Radius of Gyration, $R_g$:] The average displacement of a user from his or her center of mass: $R_g=\sqrt{\frac{1}{n}\sum_{j=1}^n|\vec{r}_j-\vec{r}_{cm}|^2}$, where $n$ is the number of calls made by a user in the month and $r_{cm}$  is the center of mass calculated by averaging the positions of all a users calls that month.  To guard against outliers such as long trips for vacation or difficulty identifying important locations due to noise, we only consider months for users where more than 5 calls were made and locations that where a user recorded more than three calls.
\item[Average Distance from Top Tower, $R_1$:] The average displacement of a user from their most called location: $R_1=\sqrt{\frac{1}{n}\sum_{j=1}^n|\vec{r}_j-\vec{r}_1|^2}$, where $n$ is the number of calls made by a user in the month and $r_1$ is the coordinates of the location most visited by the user.  To guard against outliers such as long trips for vacation or difficulty identifying important locations due to noise, we only consider months for users where more than 5 calls were made and locations that where a user recorded more than three calls.
\end{description}

\subsection{Measuring Changes}
For each user $i$, we compute these metrics monthly.  Because individuals may have different baseline behaviors, we normalize a user's time series to the month immediately before the layoff denoted $t^*$.  To assess differences in behavior as a result of the mass layoff, we construct three groups: (1) A group of laid off users from the town where the probability of being laid off is that calculated in the previous section, (2) a town control group consisting of the same users as group 1, but with inverse weights, and (3) a group of users selected at random from the country population.  Each user in the final group is weighted equally.

For each month, we compute the weighted average of all metrics then plot the difference between the laid off group and both control groups in Figure 3 in the main text.

\begin{align}
y_t &= \sum_i w_i \frac{y_{i,t}}{y_{i,t^*}} \\
\Delta y_t &= \bar{y}_t - \bar{y}_{t,control}
\end{align}
We estimate changes in monthly behavior using OLS regressions.  We specify two models that provide similar results.  For a metric :

\begin{equation}
y_i = \alpha_i + \beta_1 A_i + \beta_2 U_i + \beta_3 A_i U_i
\end{equation}
where $A_i$  is a dummy variable indicating if the observation was made in a month before or after the plant closure and $U_i$ is a dummy variable that is 1 if the user was assigned a greater than 50\% probability of having been laid off and 0 otherwise.
An alternate model substitutes the probability of layoff itself,  for the unemployed dummy:
\begin{equation}
y_i = \alpha_i+ \beta_1 A_i + \beta_2 w_i + \beta_3 A_i w_i
\end{equation}

In many cases, we are more interested in relative changes in behavior rather than absolute levels.  For this, we specify a log-level model of the form:
\begin{equation}
\log(y_i)=\alpha_i+ \phi_1 A_i + \phi_2 w_i + \phi_3 A_i w_i
\end{equation}
Now the coefficient $\phi_3$ can be interpreted as the percentage change in feature $y_{i,n}$ experienced by a laid off individual in months following the plant closure.  Changes to mobility metrics as well as changes to total, incoming, and outgoing calls were estimated using the log-level model.  Churn and To Town metrics are percentages already and are thus estimated using a level-level model.  The changes in the number of contacts each also estimated using a level-level model.

Models are estimated using data from users believed to be unemployed and data from the two control groups.  Results are shown in Table \ref{tab:micro_regression}.  Comparisons to each group produce consistent results.

\begin{figure*}
 \includegraphics[width=1\linewidth]{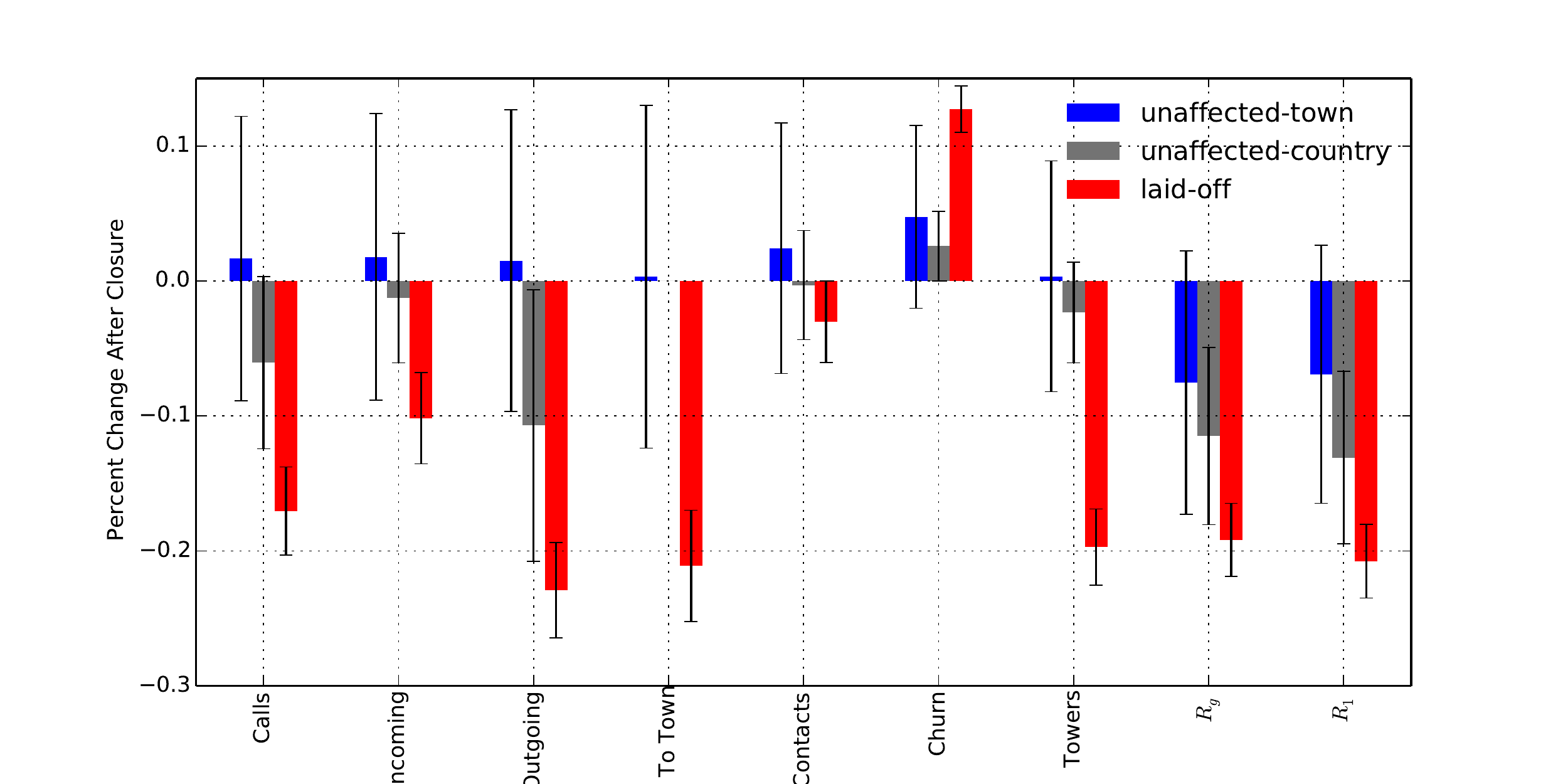}%
 \caption{\label{fig:percent_change} For each sample of mobile phone users (those we believe to be unaffected by the layoff living in the town and the country as well as those with a probability $p>0.5$ of being laid off), we plot the percent change in each variable before and after the layoff.}
 \end{figure*}

In addition relative changes, we also measure percent changes in each variable pre- and post-layoff.  Figure \ref{fig:percent_change} shows the percent change in each variable for an average month before and after the plant closing.  We report changes for three groups, those we believe were unaffected by the layoff, but live in the tower, those we believe are unaffected, but live elsewhere in the country, and those with a probability $p>0.5$ of being laid off.  The laid off group shows significant changes in all metrics, while the town and country control groups show few.

\subsection{Predicting Province Level Unemployment}
To evaluate the predictive power of micro-level behavioral changes, we use data from a different undisclosed industrialized European country.  As discussed in the main text, we use call detail records spanning nearly 3 years and the entire user base of a major mobile phone provider in the country.  For each of the roughly 50 provinces within this country, we assemble quarterly unemployment rates during the period covered by the CDR data.  At the national level, we collect a time series of GDP.  We select a sample of users in each province and measure the average relative value of 7 of the variables identified to change following a layoff.  To-town and distance from home variables are omitted as the former is only measured when we know the location of the layoff and the latter is strongly correlated with $R_g$.

First, we correlate each aggregate calling variable with unemployment at the regional level.  To control for differences in base levels of unemployment across the country, we first de-mean unemployment and each aggregate variable.   Table \ref{tab:PCA_corr} shows that each calling behavior is significantly correlated with unemployment and that these correlations are consistent with the directions found in the individual section of the paper.  Moreover, we discover strong correlation between each of the calling behavior variables, suggesting that principal component analysis is appropriate.

\subsubsection{Principal Component analysis}
As shown in the individual section of the paper, changes in these variables following a mass layoff are correlated.  This correlation is seen in province level changes as well (Table \ref{tab:PCA_corr}).  Given this correlation, we use principal component analysis (PCA) extract an independent mobile phone variable and guard against co-linearity when including all phone variables as regressors.  The results from PCA and the loadings in each component can be found in Table \ref{tab:PCA_stats} and Table \ref{tab:PCA_loadings}, respectively.  We find only the first principal component passes the Kaiser test with an eigenvalue significantly greater than 1, but that this component captures 59\% of the variance in the calling data.  The loadings in this component fall strongly on the social variables behavior.  We then compute the scores for this component for each observation in the data and use these scores as regressors.  The prominent elements of the first principal component are primarily related to the social behavior of callers.

\subsubsection{Model Specification}
We make predictions of present and future unemployment rates using three different models specifications of unemployment where each specification is run in two variants, one with the principal component score as an additional independent variable denoted as CDR$_t$ and the other without.  The twelve models are described as follows:

\begin{enumerate}
\item \textbf{AR(1)}
\begin{align}
U_t &= \alpha_1 U_{t-1} \\
U_t &= \beta_1 U_{t-1}+\gamma CDR_t \\
U_{t+1} &= \alpha_1 U_{t-1} \\
U_{t+1} &= \beta_1 U_{t-1}+\gamma CDR_t
\end{align}

\item \textbf{AR(1) + Quad}
\begin{align}
U_t &= \alpha_1 U_{t-1} + \alpha_2 U_{t-1}^2 \\
U_t &= \beta_1 U_{t-1} + \beta_2 U_{t-1}^2 +\gamma CDR_t \\
U_{t+1} &= \alpha_1 U_{t-1} + \alpha_2 U_{t-1}^2 \\
U_{t+1} &= \beta_1 U_{t-1} + \beta_2 U_{t-1}^2 +\gamma CDR_t
\end{align}

\item \textbf{AR(1) + GDP}
\begin{align}
U_t &= \alpha_1 U_{t-1} + \alpha_2 gdp_{t-1} \\
U_t &= \beta_1 U_{t-1} + \beta_2 gdp_{t-1} +\gamma CDR_t \\
U_{t+1} &= \alpha_1 U_{t-1} + \alpha_2 gdp_{t-1} \\
U_{t+1} &= \beta_1 U_{t-1} + \beta_2 gdp_{t-1} +\gamma CDR_t
\end{align}

\end{enumerate}

To evaluate the ability of these models to predict unemployment, we use a cross-validation framework.  Data from half of the provinces are used to train the model and these coefficients are used to predict unemployment rates given data for the other half of the provinces.  We perform the same procedure switching the training and testing set and combine the out of sample predictions for each case.  We evaluate the overall utility of these models by plotting predictions versus observations, finding strong correlation (see the main text).  To evaluate the additional benefit gained from the inclusion of phone data, we compute the percentage difference between the same model specification with and without the mobile phone data, $\Delta RMSE \% = 1- RMSE_{w/ CDR} / RMSE_{w/out}$ .In each case, we find that the addition of mobile phone data reduces the RMSE by 5\% to 20\%.

\subsubsection{Predictions using weekly CDR Data}
Until now, we have used data from the entire quarter to predict the results from the unemployment survey conducted in the same quarter.  While these predictions would be available at the very end of the quarter, weeks before the survey data is released, we also make predictions using CDR data from half of each quarter to provide an additional 1.5 months lead time that may increase the utility of these predictions.  We estimate the same models as described in the previous section and find similar results.  Even without full access to a quarter's CDR data, we can improve predictions of that quarter's unemployment survey before the quarter is over by 3\%-6\%.

\subsubsection{The Effect of Sample Size on Feature Estimation}
It is important to consider the extent to which the sampling size is sufficient and does not affect much the feature estimation. We study the reliability of sample size ($k$) in terms of relative standard deviation (RSD). For each given sample size $k$, we sample T times (without replacement) from the population. The RSD with respect to sample size $k$ for a particular feature, is given by $RSD(k) = \frac{s_k}{f_k}$ where $s_k$ is the standard deviation of the feature estimates from the $T$ samples, and $f_k$ is the mean of the feature estimates from the $T$ samples. We use $T=10$ to study the feature reliability. In Figure \ref{fig:RSD}, we plot the different features' \%RSD by averaging the RSD values of all provinces. The plots show that the values of \%RSD over sample size $k=100,200,..., 2000$ decrease rapidly. When sample size $k=2000$, the \%RSD for all features, except for radius of gyration ($R_g$), is lower than 1\%. The estimates of $R_g$ exhibit the highest variation; however, we can still obtain reliable estimates with thousands of sampled individuals ($RSD(k)=0.026$ for $R_g$, with $k=2000$). For the results in the manuscript, a value of k=3000 was chosen with the confidence that sample size effects are small.
 
\subsection{Mass Layoffs and General Unemployment}
While mass layoffs provide a convenient and interesting natural experiment to deploy our methods, they are only one of many employment shocks that economy absorbs each month.  We have measured changes in call behaviors due to mass layoffs, but these changes may be unhelpful if they do not result from other forms of unemployment like isolated layoffs of individual works.  Though it is beyond the scope of this work to directly determine if individuals affected by mass layoffs experience the same behavioral changes as those experiencing unemployment due to other reasons, strong correlations have been observed between the number of mass layoffs observed in a given time period and general unemployment rates - \url{http://www.bls.gov/news.release/pdf/mmls.pdf}.

\end{document}